\documentclass[11pt]{article}
\usepackage{epsfig}

\makeatletter
\newcommand\eqsecnum{
\@newctr{equation}[section]
\renewcommand\theequation{\arabic{section}.\arabic{equation}}%
}
\newcommand{\vereq}[2]{\lower3pt\vbox{\baselineskip1.5pt \lineskip1.5pt
\ialign{$\m@th#1\hfill##\hfil$\crcr#2\crcr\sim\crcr}}}
\newcommand{\lesssim}{\mathrel{\mathpalette\vereq<}}

\renewcommand\appendix{\par
  \setcounter{section}{0}%
  \setcounter{subsection}{0}%
  \gdef\thesection{\@Alph\c@section}%
\renewcommand\theequation{\Alph{section}.\arabic{equation}}
}
\newcount\@indentflag \global\@indentflag=1 %
\newdimen\@eqtoeqnum \@eqtoeqnum=6pt %
\def\@indentamount{%
\ifcase\@indentflag 0pt\or\@centering\or0pt plus1fil\fi\relax
}
\def\FL{\global\@indentflag=0 }
\def\FR{\global\@indentflag=2 }
\def\@eqnnum{\hbox{\reset@font\rm(\theequation)}}
\let\make@eqnnum=\@eqnnum %
\def\eqnum#1{\dec@eqnnum \global\def\make@eqnnum{\reset@font\rm(#1)}%
\def\@currentlabel{#1}%
}
\def\inc@eqnnum{\addtocounter{equation}{1}}
\def\dec@eqnnum{\addtocounter{equation}{-1}}
\newbox\@testboxa
\newbox\@testboxb
\def\equation{\par\vskip-\lastskip\vskip\abovedisplayskip
\inc@eqnnum\let\@currentlabel=\theequation
\setbox\@testboxa=\hbox\bgroup\hskip\@totalleftmargin\hskip\@indentamount
\hbox\bgroup$\displaystyle
}
\def\endequation{$\egroup\hskip\@centering\egroup %
\setbox\@testboxb=\hbox{\make@eqnnum}%
\bgroup
\@tempdima\wd\@testboxa \advance\@tempdima by\wd\@testboxb
\ifcase\@indentflag
\advance\@tempdima by\@eqtoeqnum
\ifdim\@tempdima<\hsize %
\def\@tempa{0}%
\else
\def\@tempa{1}%
\fi
\or
\advance\@tempdima by2\@eqtoeqnum
\ifdim\@tempdima<\hsize %
\def\@tempa{0}%
\else %
\@tempdima\wd\@testboxa \advance\@tempdima by\wd\@testboxb
\advance\@tempdima by\@eqtoeqnum
\ifdim\@tempdima<\hsize %
\def\@tempa{0}%
\setbox\@testboxa\hbox{\hfill\box\@testboxa\kern\@eqtoeqnum}%
\else
\def\@tempa{1}%
\fi
\fi
\or
\advance\@tempdima by2\@eqtoeqnum
\ifdim\@tempdima<\hsize %
\def\@tempa{0}%
\setbox\@testboxb=\hbox{\kern\@eqtoeqnum\make@eqnnum}%
\else
\def\@tempa{1}%
\fi
\fi
\ifnum\@tempa=0 %
\hbox to\hsize{\unhbox\@testboxa\box\@testboxb}%
\else %
\vbox{\hbox to\hsize{\unhbox\@testboxa}%
\vskip6pt %
\hbox to\hsize{\hfil\box\@testboxb}}%
\fi
\egroup
\global\let\make@eqnnum\@eqnnum %
\vskip\belowdisplayskip\noindent\global\@indentflag=1 \global\@ignoretrue
}
\def\eqnarray{\par\vskip-\lastskip\vskip\abovedisplayskip
\inc@eqnnum\let\@currentlabel=\theequation
\global\@eqnswtrue\m@th
\global\@eqcnt\z@
\tabskip\@totalleftmargin\advance\tabskip by\@indentamount\let\\\@eqncr
\halign to\hsize\bgroup\hskip\@centering
$\displaystyle\tabskip\z@{##{}}$&\global\@eqcnt\@ne
\hfil${{}##{}}$\hfil
&\global\@eqcnt\tw@ $\displaystyle\tabskip\z@{##}$\hfil
\tabskip\@centering \if@eqnsw\phantom{\make@eqnnum\kern\@eqtoeqnum}\fi
&\llap{##}\tabskip\z@\cr}
\def\endeqnarray{%
\@@eqncr\egroup
\vskip\belowdisplayskip\noindent
\dec@eqnnum\global\@indentflag=1
\global\let\make@eqnnum\@eqnnum %
\global\@ignoretrue
}
\def\be{\begin{eqnarray}}\def\ee{\end{eqnarray}}
\def\bi{\bibitem}

\def\la{\langle}\def\ra{\rangle}\def\Tr{{\rm Tr}}
\def\L{{\cal L}}
\def\del{\partial}
\def\roughly#1{\mathrel{\raise.3ex\hbox{$#1$\kern-.75em%
\lower1ex\hbox{$\sim$}}}}

\renewcommand{\thefootnote}{\fnsymbol{footnote}}

\makeatother

\eqsecnum

\begin{document}

\begin{titlepage}

\begin{flushright}
\begin{minipage}{3cm}
\begin{flushleft}
SNUTP 02/021\\
DPNU-02-19
\end{flushleft}
\end{minipage}
\end{flushright}

\begin{center}
{\Large\bf Effective Degrees of Freedom at Chiral Restoration\\
and the Vector Manifestation in HLS theory}
\end{center}
\vspace{1cm plus 0.5cm minus 0.5cm}
\begin{center}
\large Masayasu Harada$^{(a)}$~\footnote{%
  Present address:
  {\it Department of Physics, Nagoya University, Nagoya 464-8602,
  Japan.}
},
Youngman Kim$^{(a)}$~\footnote{and
\it Department of Physics and Astronomy, University of South
Carolina, Columbia, SC 29208},
 Mannque  Rho$^{(b,c)}$ \\
and Chihiro Sasaki$^{(d)}$
\end{center}
\vspace{0.5cm plus 0.5cm minus 0.5cm}
\begin{center}
(a)~{\it  School of Physics, Seoul National University,
Seoul 151-742, Korea}\\
 (b)~{\it Service de Physique Th\'eorique, CEA/DSM/SPhT,
Unit\'e de recherche associ\'ee au CNRS,
CEA/Saclay,  91191 Gif-sur-Yvette c\'edex, France}\\
(c)~{\it School of Physics, Korea Institute for Advanced Study,
Seoul 130-722, Korea}\\
(d)~{\it Department of Physics, Nagoya University, Nagoya,
464-8602, Japan.}
\end{center}
\vspace{0.6cm plus 0.5cm minus 0.5cm}

\begin{abstract}
The question as to what the relevant {\it effective} degrees of
freedom at the chiral phase transition are remains largely
unanswered and must be addressed in confronting both terrestrial
and space laboratory observations purporting to probe matter under
extreme conditions. We address this question in terms of the
vector susceptibility $\chi_V$ (VSUS in short) and the
axial-vector susceptibility $\chi_A$ (ASUS in short) at the
temperature-induced chiral transition. We consider two possible,
albeit simplified, cases that are contrasting, one that is given
by the standard chiral theory where only the pions  
figure in the vicinity
of the transition and the other that is described by hidden local
symmetry (HLS) theory with the Harada-Yamawaki vector
manifestation (VM) where nearly massless vector mesons also enter.
We find that while in the standard chiral theory, the pion velocity
$v_\pi$ proportional to the ratio of the space component $f_\pi^s$
of the pion decay constant over the time component $f_\pi^t$ tends
to zero near chiral restoration with $f_\pi^t\neq 0$, in the
presence of the vector mesons with vanishing mass, the result is
drastically different: HLS with VM {\it predicts} that $\chi_V$
automatically equals $\chi_A$ in consistency with chiral
invariance and that $v_\pi\sim 1$ with $f_\pi^t\approx
f_\pi^s\rightarrow 0$ as $T\rightarrow T_c$. These results are
obtained in the leading order in power counting but we expect
their qualitative features to remain valid more generally in the
chiral limit {\it thanks to the VM point}.
\end{abstract}

\end{titlepage}

\newpage
\renewcommand{\thefootnote}{\#\arabic{footnote}}
\setcounter{footnote}{0}
\section{Introduction}
\indent\indent One of the most crucial questions to answer in the
effort to understand chiral restoration in relativistic heavy-ion
collisions as well as in dense medium as in compact stars is: What
are the relevant {\it effective} degrees of freedom just before
and after the phase transition? The standard scenario, generally
accepted in the community, is that the only relevant excitations
in the broken symmetry sector near the phase transition are the
pions, the pseudo-Nambu-Goldstone modes of broken chiral symmetry,
i.e.,  the standard chiral theory. However there is no a 
priori reason to exclude other scenarios. In fact, it has been
argued by Harada and Yamawaki~\cite{HY:VM,HY:PR} that
 the vector manifestation (VM) with the massless (in the
chiral limit) vector mesons can
 correctly describe chiral restoration.
In a recent attempt to understand some of the puzzling
results coming out of relativistic heavy ion experiments at RHIC,
Brown and Rho~\cite{QM2002} invoked the VM scenario in which the
``light" $\rho$ mesons play a crucial role: The vector mesons
there are ``relayed" via a Higgsing to the gluons in the QCD
sector.

In this paper, we address the issue of what the relevant degrees
of freedom can be at the chiral transition induced by high
temperature and their possible implications on observables in
heavy-ion physics. In doing this, we focus on the vector and
axial-vector susceptibilities and the pion velocity very near the
critical temperature $T_c$ using the result obtained in
\cite{HaradaSasaki} who have shown that the VM holds at $T=T_c$.
The issue of what happens at high density is discussed in
\cite{LPRV}.

As a way of introduction to the main objective of this paper, we
begin by briefly summarizing the arguments and results obtained in
the standard chiral theory scenario~\cite{SS:1,SS:2}.

Consider the vector isospin susceptibility (VSUS) $\chi_V$
(denoted by SS as $\chi_{I}$) and the axial-vector isospin
susceptibility (ASUS) $\chi_A$ (denoted by SS as $\chi_{I5}$)
defined in terms of the vector charge density $V_0^a (x)$ and the
axial-vector charge density $A_0^a (x)$ by the Euclidean
correlators:
 \be
\delta^{ab}\chi_V&=&
\int^{1/T}_0 d\tau\int d^3\vec{x}\la V_0^a
(\tau, \vec{x}) V_0^b (0,\vec{0})\ra_\beta,\\
\delta^{ab}\chi_A&=& \int^{1/T}_0 d\tau\int d^3\vec{x}\la A_0^a
(\tau, \vec{x}) A_0^b (0,\vec{0})\ra_\beta
 \ee
where $\la ~\ra_\beta$ denotes thermal average and
 \be
V_0^a\equiv \bar{\psi}\gamma^0\frac{\tau^a}{2}\psi, \ \
A_0^a\equiv \bar{\psi}\gamma^0\gamma^5\frac{\tau^a}{2}\psi
 \ee
with the quark field $\psi$ and the $\tau^a$ Pauli matrix the
generator of the flavor $SU(2)$.

We are interested in these SUS's near the critical temperature
$T=T_c$ at zero baryon density $n=0$. In particular we would like
to compute them ``bottom-up" approaching $T_c$ from below. In
order to do this, we need to resort to effective field theory of
QCD which requires identifying, in the premise of an EFT, {\it all}
the relevant degrees of freedom.

Let us first assume as done by Son and Stephanov
(SS)~\cite{SS:1,SS:2} that the only relevant effective degrees of
freedom in heat bath are the pions, and that all other degrees of
freedom can be integrated out with their effects incorporated into
the coefficients of higher order terms in the effective
Lagrangian. Here the basic assumption is that near chiral
restoration, there is no instability in the channel of the degrees
of freedom that have been integrated out. In this pion-only case,
the appropriate effective Lagrangian for the axial correlators is
the in-medium chiral Lagrangian dominated by the current algebra
terms,
 \be
\L_{eff}=\frac{{f_\pi^t}^2}{4}\left(\Tr\nabla_0 U\nabla_0
U^\dagger - v_\pi^2\Tr\del_i U\del_i U^\dagger\right) -\frac 12
\la\bar{\psi}\psi\ra {\rm Re} M^\dagger
U\label{LA}+\cdots\label{Leff}
 \ee
where $v_\pi$ is the pion velocity, $M$ is the mass matrix
introduced as an external field, $U$ is the chiral field and the
covariant derivative $\nabla_0 U$ is given by $\nabla_0 U=\del_0 U
-\frac i2 \mu_A (\tau_3 U +U\tau_3)$ with $\mu_A$ the axial
isospin chemical potential. The ellipsis stands for higher order
terms in spatial derivatives and covariant
derivatives.~\footnote{The notation here deviates a bit from that
of SS. For example, it will turn out that the pion velocity will
have the form $v_\pi^2=f_\pi^s/f_\pi^t$ (see Eq.(\ref{v2 rel}))
where $f_\pi^t$ ($f_\pi^s$) is the temporal (spatial) component of
the pion decay constant.} Given the effective action described by
(\ref{Leff}) with possible non-local terms ignored, then the ASUS
takes the simple form
 \be
\chi_A=-\frac{\del^2}{\del\mu_A^2}\L_{eff}|_{\mu_A=0}={f_\pi^t}^2.
 \ee
The principal point to note here is that {\it as long as the
effective action is given by local terms (subsumed in the
ellipsis) involving the $U$ field, this is the whole story}: There
is no other contribution to the ASUS than the temporal component
of the pion decay constant.

Next one assumes that at the chiral phase transition point
$T=T_c$, the restoration of chiral symmetry dictates the equality
 \be
\chi_A=\chi_V.
 \ee
While there is no lattice information on $\chi_A$, $\chi_V$ has
been measured as a function of
temperature~\cite{GLTRS,Brown-Rho:96}. In particular, it is
established that
 \be
\chi_V|_{T=T_c}\neq 0,
 \ee
which leads to the conclusion~\cite{SS:1,SS:2} that
 \be
f_\pi^t|_{T=T_c}\neq 0.
 \ee
On the other hand, it is expected and verified by lattice
simulations that the space component of the pion decay constant
$f_\pi^s$ should vanish at $T=T_c$. One therefore arrives at
 \be
v_\pi^2\sim f_\pi^s/f_\pi^t\rightarrow 0, \ \ T\rightarrow T_c.
 \ee
This is the main conclusion of the standard chiral theory.

To check whether this prediction is firm, let us see what one
obtains for the VSUS in the same effective field theory approach.
The effective Lagrangian for calculating the vector correlators is
of the same form as the ASUS, Eq.~(\ref{Leff}), except that the
covariant derivative is now defined with the vector isospin
chemical potential $\mu_V$ as $\nabla_0 U=\del_0 U-\frac 12 \mu_V
(\tau_3 U-U\tau_3)$. Now if one assumes as done above for $\chi_A$
that possible non-local terms can be dropped, then the SUS is
given  by
 \be
\chi_V=-\frac{\del^2}{\del\mu_V^2}\L_{eff}|_{\mu_V=0}
 \ee
which can be easily evaluated from the Lagrangian. One finds that
 \be
\chi_V=0
 \ee
{\it for all temperature.} While it is expected to be zero at
$T=0$, the vanishing $\chi_V$ for $T\neq 0$ is at variance with
the lattice data at $T=T_c$.~\footnote{%
  The reason for this defect is
  explained in terms of hydrodynamics by Son and
  Stephanov~\cite{SS:1,SS:2}.
}

We now turn to the main objective of this paper: the prediction by
the vector manifestation (VM)~\cite{HY:VM}.
Basically
the same scenario was suggested some time ago in conjunction with
Brown-Rho scaling~\cite{Brown-Rho:91,Brown-Rho:01b}. As will be
shown in detail in the following sections, the VM {\it requires} that
the vector mesons
figure on the same footing
with the pions as {\it the} relevant degrees of freedom as the chiral
transition point is approached from below. The key reason for this
conclusion is that the chiral transition coincides with the VM
fixed point at which the vector meson mass
must vanish in the chiral limit~\cite{HY:fate}.
This means that the
vector-meson degrees of freedom
{\it cannot} be integrated out near chiral restoration.

Our principal results - which are basically different from the
standard chiral theory scenario - can be summarized as follows. In the presence
of the $\rho$-meson, the only approach that is consistent with
chiral perturbation theory is the hidden local symmetry (HLS) with
the VM fixed point~\footnote{It has been stressed in the
literature (see, e.g., \cite{Georgi,HY:PR}) -- and is stressed
again -- that HLS is a bona-fide effective field theory of QCD
{\it only} if the $\rho$-meson mass is considered as of the same
chiral order as the pion mass. In HLS theory, this condition is
naturally met by the $\rho$-meson mass near the chiral transition
point, so chiral perturbation theory should be more effective in
this regime. This point that underlines our arguments that follow
justifies our one-loop calculation.}. The present analysis based
on this theory predicts~\footnote{The reason for that the $v_\pi$
deviates from 1 will be explained below.}
 \be
f_\pi^t \vert_{T=T_c}=
f_\pi^s \vert_{T=T_c}=0 \ , \qquad
\left. v_\pi\right\vert_{T=T_c} \lesssim 1
\label{main1}
  \ee
and
\begin{equation}
\chi_A|_{T=T_c}=\chi_V|_{T=T_c}=
\frac{N_f^2}{6} T_c^2
\ ,\label{main2}
\end{equation}
where we have included the normalization factor of $2N_f$. Note
that the equality of $\chi_A$ and $\chi_V$ at $T=T_c$ is an output
of the theory. This result is a direct consequence of the fact
that the $\rho$ and $\pi$
enter on the same
footing in the VM:
At the VM fixed point, the longitudinal components of the
vector mesons and the pions form a degenerate multiplet.

The rest of the paper is devoted to the derivation of the main
results (\ref{main1}) and (\ref{main2}). In Section~\ref{sec:HLS},
hidden
local symmetry (HLS) theory
is briefly introduced.
Section~\ref{sec:TPF} describes how
thermal two-point functions are calculated in the HLS theory.
In
Section~\ref{sec:CC},
we write down the in-medium vector and axial-vector
current correlators that are needed in what follows. Pion decay
constants and pion velocity are computed in the given framework in
Section~\ref{sec:PDCV}.
The susceptibilities are defined in Section~\ref{sec:SUS} and
computed for temperature $T\sim T_c$.
The conclusion is given in Section~\ref{sec:SR}.
The Appendices contain explicit formulas used in the main text.
A more extensive treatment of the material covered in this paper
together with other issues of finite temperature effective field
theory in the VM is found in Ref.~\cite{HaradaSasaki:prep}.

\section{Hidden Local Symmetry}
\label{sec:HLS} \indent\indent In this Section, we briefly
summarize the HLS model. Our discussion will be highly sketchy.
For details, the readers are invited to the review~\cite{HY:PR}.
As mentioned above, the HLS spin-1 field is assumed to be as
relevant as the pion field near chiral restoration.
The HLS model~\cite{BKUYY,BKY:88} is based on the $G_{\rm
global} \times H_{\rm local}$ symmetry, where $G =
\mbox{SU($N_f$)}_{\rm L} \times \mbox{SU($N_f$)}_{\rm R}$  is the
global chiral symmetry and $H = \mbox{SU($N_f$)}_{\rm V}$ is the
HLS. The basic quantities are the HLS gauge field $V_\mu$ and two
variables or ``coordinates"
\begin{eqnarray}
&&
\xi_{\rm L,R} = e^{i\sigma/F_\sigma} e^{\mp i\pi/F_\pi}
\ ,
\end{eqnarray}
where $\pi$ denotes the pseudoscalar Nambu-Goldstone (NG) boson
and $\sigma$ the NG boson absorbed into the HLS gauge field
$V_\mu$ (longitudinal $\rho$). $F_\pi$ and $F_\sigma$ are
corresponding decay constants, and the parameter $a$ is defined as
$a \equiv F_\sigma^2/F_\pi^2$. The transformation property of
$\xi_{\rm L,R}$ is given by
\begin{eqnarray}
&&
\xi_{\rm L,R}(x) \rightarrow \xi_{\rm L,R}^{\prime}(x) =
h(x) \xi_{\rm L,R}(x) g^{\dag}_{\rm L,R}
\ ,
\end{eqnarray}
where $h(x) \in H_{\rm local}$ and
$g_{\rm L,R} \in G_{\rm global}$.
The covariant derivatives of $\xi_{\rm L,R}$ are defined by
\begin{eqnarray}
&&
D_\mu \xi_{\rm L} =
\partial_\mu \xi_{\rm L} - i V_\mu \xi_{\rm L}
+ i \xi_{\rm L} {\cal L}_\mu
\ ,
\nonumber\\
&&
D_\mu \xi_{\rm R} =
\partial_\mu \xi_{\rm R} - i V_\mu \xi_{\rm R}
+ i \xi_{\rm R} {\cal R}_\mu
\ ,
\label{covder}
\end{eqnarray}
where
${\cal L}_\mu$ and ${\cal R}_\mu$ denote the external gauge fields
gauging the $G_{\rm global}$ symmetry.
{}From the above covariant derivatives two 1-forms
are constructed as
\begin{eqnarray}
&&
\hat{\alpha}_{\perp}^\mu =
( D_\mu \xi_{\rm R} \cdot \xi_{\rm R}^\dag -
  D_\mu \xi_{\rm L} \cdot \xi_{\rm L}^\dag
) / (2i)
\ ,
\nonumber\\
&&
\hat{\alpha}_{\parallel}^\mu =
( D_\mu \xi_{\rm R} \cdot \xi_{\rm R}^\dag +
  D_\mu \xi_{\rm L} \cdot \xi_{\rm L}^\dag
) / (2i)
\ .
\end{eqnarray}

It should be noticed that,
as first pointed by Georgi in Ref.~\cite{Georgi}
and developed further in Refs.~\cite{Tanabashi,HY:WM,HY:PR},
the systematic chiral perturbation can be performed with
including the vector meson loop in addition to the pion loop
in the HLS.
The expansion parameter is a ratio of the $\rho$
meson mass to the chiral symmetry breaking scale
$\Lambda_\chi$, $m_\rho/\Lambda_\chi$, in addition to the ratio of the
momentum $p$ to $\Lambda_\chi$, $p/\Lambda_\chi$,
as used in the ordinary chiral perturbation theory.
The counting scheme is made as in the ordinary chiral perturbation
theory by assigning ${\cal O}(p)$ to the HLS gauge coupling
$g$~\cite{Georgi,Tanabashi}.

With the above counting scheme
the Lagrangian at the leading order, counted as ${\cal O}(p^2)$,
is given
by~\cite{BKUYY,BKY:88}
\begin{equation}
{\cal L} = F_\pi^2 \, \mbox{tr}
\left[ \hat{\alpha}_{\perp\mu} \hat{\alpha}_{\perp}^\mu \right]
+ F_\sigma^2 \, \mbox{tr}
\left[
  \hat{\alpha}_{\parallel\mu} \hat{\alpha}_{\parallel}^\mu
\right]
- \frac{1}{2g^2} \mbox{tr} \left[ V_{\mu\nu} V^{\mu\nu} \right]
\ ,
\label{Lagrangian}
\end{equation}
where $g$ is the HLS gauge coupling and
$V_{\mu\nu} = \partial_\mu V_\nu - \partial_\nu V_\mu
- i [ V_\mu , V_\nu ]$ the gauge field strength.
When the kinetic term of the gauge field
is ignored in the low-energy region,
the second term of Eq.(\ref{Lagrangian}) vanishes
by integrating out $V_\mu$
and only the first term remains.
Then,
the HLS model is reduced to the nonlinear sigma model based on $G/H$.

The one-loop quantum corrections
calculated from the leading order Lagrangian in Eq.~(\ref{Lagrangian})
are counted as ${\cal O}(p^4)$.
The divergences generated at ${\cal O}(p^4)$ are renormalized by
the ${\cal O}(p^4)$ terms, a complete list of which are given
in Refs.~\cite{Tanabashi,HY:PR}.
Here we show the
terms of the ${\cal O}(p^4)$ Lagrangian relevant to the present
analysis~\cite{Tanabashi,HY:WM}:
\begin{equation}
  {\cal{L}}_{(4)} = z_1\mbox{tr}\bigl[ \hat{\cal{V}}_{\mu\nu}
                       \hat{\cal{V}}^{\mu\nu} \bigr] +
                    z_2\mbox{tr}\bigl[ \hat{\cal{A}}_{\mu\nu}
                       \hat{\cal{A}}^{\mu\nu} \bigr] +
                    z_3\mbox{tr}\bigl[ \hat{\cal{V}}_{\mu\nu}
                       V^{\mu\nu} \bigr], \label{eq:L(4)}
\end{equation}
where
 \begin{eqnarray}
  \hat{\cal{A}}_{\mu\nu}=\frac{1}{2}
                         \bigl[ \xi_R{\cal{R}}_{\mu\nu}\xi_R^{\dagger}-
                                \xi_L{\cal{L}}_{\mu\nu}\xi_L^{\dagger}
                         \bigr]\ ,
\nonumber\\
  \hat{\cal{V}}_{\mu\nu}=\frac{1}{2}
                         \bigl[ \xi_R{\cal{R}}_{\mu\nu}\xi_R^{\dagger}+
                                \xi_L{\cal{L}}_{\mu\nu}\xi_L^{\dagger}
                         \bigr]\ ,
  \label{def V mn}
\end{eqnarray}
with ${\cal{R}}_{\mu\nu}\ \mbox{and}\ {\cal{L}}_{\mu\nu}$ being
the field strengths of ${\cal{R}}_{\mu}\ \mbox{and}\ {\cal{L}}_{\mu}$.

We should stress that
we assume that we obtain the bare HLS Lagrangian by
integrating out the quark and gluon degrees of freedom
at the
matching scale $\Lambda$, so that
bare parameters of the above HLS Lagrangian
such as $F_{\pi\,,{\rm bare}}$
defined at $\Lambda$
are determined through the Wilsonian
matching between the HLS and the underlying QCD~\cite{HY:WM}:
As we briefly review in Appendix~\ref{ssec:WMCT0},
bare parameters are determined through the Wilsonian matching
conditions (\ref{match A})--(\ref{match z}) obtained by matching
the axial-vector and vector current correlators in the HLS
with those in the operator product expansion (OPE).
As was shown in Refs.~\cite{HY:WM,HY:PR},
these bare parameters are scaled down to the low energy region
through the Wilsonian renormalization group equations (RGEs)
to predict
several physical quantities in remarkable agreement with
experiment.

Let us extend the above HLS Lagrangian to the analysis
in hot matter.
We assume that we obtain the bare HLS Lagrangian by integrating
out the quarks and gluons degrees of freedom at the matching scale
$\Lambda$ {\it in the presence of medium},
and then the bare parameters are determined by matching
the HLS to the underlying QCD.
As was shown in Ref.~\cite{HaradaSasaki} and briefly
reviewed in Appendix~\ref{ssec:ITDBP},
when we make the matching in the presence of hot matter,
the bare parameters have the {\it intrinsic temperature dependences}.
In general, the Lorentz non-scalar operators such as
$\bar{q}\gamma_\mu D_\nu q$ exist in
the form of the current correlators derived by the
OPE~\cite{Hatsuda-Koike-Lee}.
However, as discussed in Appendix~\ref{ssec:ITDBP},
such Lorentz symmetry violating contributions
caused from the Lorentz non-scalar operators
are suppressed by, at least, a factor of $1/\Lambda^6$
compared with $1 + \alpha_s/\pi$,
and
the
Lorentz violating effects in the bare $\pi$ decay constant
and the bare $\sigma$ (longitudinal $\rho$) decay constant
are small:
At bare level, the difference between $F_{\pi,{\rm bare}}^t$
and $F_{\pi,{\rm bare}}^s$ as well as that between
$F_{\sigma,{\rm bare}}^t$ and $F_{\sigma,{\rm bare}}^s$
is small.
Furthermore, since we will study the physical pion decay
constants and the vector and axial-vector susceptibilities
only near the critical temperature,
the Lorentz violating effect in the HLS gauge coupling,
which may distinguish
$g_T$ from $g_L$ (see Appendix~\ref{ssec:ITDBP}),
is irrelevant
due to the decoupling nature of the transverse $\rho$
near the critical temperature in the VM.
Thus, here we use the
Lagrangian (\ref{Lagrangian})
with Lorentz invariance even at non-zero temperature to calculate
the quantum and hadronic thermal corrections.
We should stress that the
{\it explicit} Lorentz violation in medium
-- which is not negligible -- is of course taken into account (see
Appendices).

\section{Two-Point Functions}
\label{sec:TPF}

 \indent\indent We need to consider two-point
functions involving the isovector vector and axial-vector
currents.
Note that, when we calculate the hadronic thermal corrections,
we assigned ${\cal O}(p)$ to the temperature $T$ as in the
approach based on the ordinary chiral perturbation
theory~\cite{Gasser-Leutwyler}.~\footnote{%
  The same treatment within the framework of the HLS was done
  before in Refs.~\cite{Harada-Shibata,HaradaSasaki}.
}
Let us calculate them to one-loop order as in
Refs.~\cite{HY:WM,HY:PR}.
In calculating the loops, we adopt the background
field gauge (see Refs.~\cite{HY:WM,HY:PR} for details in HLS
theory) and the imaginary time formalism (see, e.g.,
Ref.~\cite{Kap}). For convenience, we introduce the following
Feynman integrals to calculate the one-loop hadronic thermal
corrections and quantum corrections to the two-point functions:
\begin{eqnarray}
A_{0}(M;T) &\equiv&
T \sum_{n=-\infty}^{\infty}
\int \frac{d^3k}{(2\pi)^3}
\frac{1}{M^2-k^2}
\ ,
\label{def:A0 2}
\\
B_{0}(p_0,\bar{p};M_1,M_2;T) &\equiv&
T \sum_{n=-\infty}^{\infty}
\int \frac{d^3k}{(2\pi)^3}
\frac{1}{ [M_1^2-k^2] [M_2^2-(k-p)^2] }
\ ,
\label{def:B0 2}
\\
B_{}^{\mu\nu}(p_0,\vec{p};M_1,M_2;T) &\equiv&
T \sum_{n=-\infty}^{\infty}
\int \frac{d^3k}{(2\pi)^3}
\frac{\left(2k-p\right)^\mu \left(2k-p\right)^\nu}{%
 [M_1^2-k^2] [M_2^2-(k-p)^2] }
\ ,
\label{def:Bmunu 2}
\end{eqnarray}
where $\bar{p}\equiv\vert\vec{p}\vert$, and
the $0$th component of the loop momentum is taken
as $k^0 = i 2 n \pi T$,
while that of the external momentum is taken as $p^0 = i 2
n^\prime \pi T$ ($n^\prime$: integer). Using the standard formula
(see, e.g., Ref.~\cite{Kap}), we can convert the Matsubara
frequency sum into an integral over $k_0$ with $k_0$ taken as the
zeroth component of a Minkowski four vector. Accordingly, the
above functions are divided into two parts as
\begin{eqnarray}
A_{0}(M;T) &=&
A_{0}^{\rm{(vac)}}(M) + \bar{A}_{0}(M;T) \ ,
\nonumber\\
B_{0}(p_0,\bar{p};M_1,M_2;T) &=&
B_{0}^{\rm{(vac)}}(p_0,\bar{p};M_1,M_2) +
\bar{B}_{0}(p_0,\bar{p};M_1,M_2;T) \ ,
\nonumber\\
B^{\mu\nu}(p_0,\vec{p};M_1,M_2;T)
&=&
B^{\rm{(vac)}\mu\nu}(p_0,\vec{p};M_1,M_2)
+
\bar{B}^{\mu\nu}(p_0,\vec{p};M_1,M_2;T)
\ ,
\label{B bar defs}
\end{eqnarray}
where $A_{0}^{\rm{(vac)}}$, $B_{0}^{\rm{(vac)}}$ and
$B^{\rm{(vac)}\mu\nu}$ are given by replacing $T
\sum_{n=-\infty}^{\infty}$ with $\int \frac{d k_0}{2\pi i}$ in
Eqs.~(\ref{def:A0 2})--(\ref{def:Bmunu 2}), and
$\bar{A}_{0}$, $\bar{B}_{0}$ and
$\bar{B}^{\mu\nu}$ are defined by Eq.~(\ref{B bar defs}).
In the present analysis, the
forms of $A_{0}^{\rm{(vac)}}$, $B_{0}^{\rm{(vac)}}$ and
$B^{\rm{(vac)}\mu\nu}$ are equivalent to the zero-temperature
ones. Then, with $p_0$ taken as the 0th component of the Minkowski
four vector, they have no {\it explicit} temperature dependence while
the intrinsic dependence mentioned above remains. Therefore, the
functions $A_{0}^{\rm{(vac)}}$, $B_{0}^{\rm{(vac)}}$ and
$B^{\rm{(vac)}\mu\nu}$ represent quantum corrections. In
$\bar{B}_{0}$ and $\bar{B}^{\mu\nu}$ one can perform
the analytic continuation of $p_0$ to the Minkowski variable after
integrating over $k_0$: Here $p_0$ is understood as $p_0 +
i\epsilon$ ($\epsilon \rightarrow +0$) for the retarded function
and $p_0 - i \epsilon$ for the advanced function. It should be
noticed that the argument $T$ in the above functions refers to
only the temperature dependence arising from the hadronic thermal
effects and {\it not} to the intrinsic thermal effects included in the
parameters of the Lagrangian.

\begin{figure}
\begin{center}
\epsfxsize = 14cm
\ \epsfbox{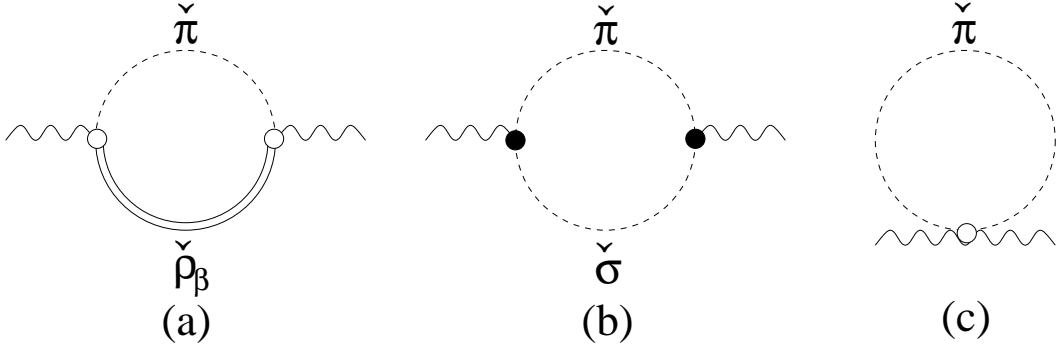}
\end{center}
\caption[]{%
Feynman diagrams contributing to the $\overline{\cal
A}_\mu$-$\overline{\cal A}_\nu$ two-point function. Here
$\check{\pi}$ represents the quantum pion field and likewise for
the others.}\label{fig:AA}
\end{figure}

Now, let us calculate the one-loop corrections to the two-point
function of the axial-vector (background) field $\overline{\cal
A}^\mu$. This is obtained by the sum of one particle irreducible
diagrams with two legs of the axial-vector background field
$\overline{\cal A}^\mu$. In Fig.~\ref{fig:AA} we show the Feynman
diagrams contributing to the $\overline{\cal
A}_\mu$-$\overline{\cal A}_\nu$ two-point function at one-loop
level.
With the help of (\ref{B bar defs}), one can express the one-loop
corrections to the two-point function in a simple form.
The corrections from $\rho$ and/or $\pi$
shown in Figs.~\ref{fig:AA}(a)--(c)
lead to
the two-point function
\begin{eqnarray}
  \Pi_\perp^{\mbox{\scriptsize(1-loop)}\mu\nu}(p_0,\vec{p};T)
&=&
- N_f a M_\rho^2 g^{\mu\nu} B_{0}(p_0,\bar{p};M_\rho,0;T)
{}+ N_f \frac{a}{4} B^{\mu\nu}(p_0,\vec{p};M_\rho,0;T)
\nonumber\\
&&
{}+N_f(a-1) g^{\mu\nu} A_{0}(0,T)
\ ,
\label{PiA boson}
\end{eqnarray}
where $B_{0}$-term comes from Fig.~\ref{fig:AA}(a), and
$B^{\mu\nu}$-term and $A_{0}$-term from
Fig.~\ref{fig:AA}(b) and Fig.~\ref{fig:AA}(c), respectively.
Combining the above loop corrections with
the tree contribution given by
\begin{eqnarray}
\Pi_\perp^{{\rm(tree)}\mu\nu}(p_0,\vec{p})
= g^{\mu\nu} F_{\pi,{\rm bare}}^2 +
2 z_{2,{\rm bare}} (g^{\mu\nu} p^2 - p^\mu p^\nu )
\ ,
\label{Pi A tree}
\end{eqnarray}
we have the two-point function of $\overline{\cal
A}_\mu$-$\overline{\cal A}_\nu$ at one-loop level as
\begin{equation}
\Pi_\perp^{\mu\nu}(p_0,\vec{p};T)
= \Pi_\perp^{{\rm(tree)}\mu\nu}(p_0,\vec{p})
+ \Pi_\perp^{\mbox{\scriptsize(1-loop)}\mu\nu}(p_0,\vec{p};T)
\ .
\end{equation}
Analogously to Eq.~(\ref{B bar defs}), we split the two-point
function into two parts as
\begin{equation}
\Pi_\perp^{\mu\nu}(p_0,\bar{p};T)
= \Pi_\perp^{{\rm(vac)}\mu\nu}(p_0,\bar{p})
+ \bar{\Pi}_\perp^{\mu\nu}(p_0,\bar{p};T)
\ ,
\label{Pi A div}
\end{equation}
where $\Pi_\perp^{{\rm(vac)}\mu\nu}$ includes the quantum
correction and the contribution at tree level in Eq.~(\ref{Pi A
tree}), and $\bar{\Pi}_\perp^{\mu\nu}$ represents the hadronic
thermal correction. Since the hadronic thermal correction
$\bar{\Pi}_\perp^{\mu\nu}$ has no divergences, the renormalization
conditions for $F_\pi^2$ and $z_2$ can be determined from
$\Pi_\perp^{{\rm(vac)}\mu\nu}$. For $F_\pi^2$ we adopt the
``on-shell" renormalization condition:
\begin{equation}
\Pi_\perp^{{\rm(vac)}\mu\nu}(p_0=0,\vec{p}=\vec{0})
= g^{\mu\nu} F_\pi^2(0)
\ .
\end{equation}
{}From this renormalization condition, we obtain the
$g^{\mu\nu}$-part of $\Pi_\perp^{{\rm(vac)}\mu\nu}$ in the
form~\cite{HY:PR}
\begin{equation}
p_\mu p_\nu \Pi_\perp^{{\rm(vac)}\mu\nu}(p_0,\vec{p})
= p^2 \left[
  F_\pi^2(0) + \widetilde{\Pi}_\perp^S(p^2)
\right]
\ ,
\end{equation}
where $\widetilde{\Pi}_\perp^S(p^2)$ is the finite renormalization
contribution satisfying
\begin{equation}
\widetilde{\Pi}_\perp^S(p^2=0) = 0 \ .
\end{equation}
For $z_2$ we adopt the renormalization condition that
$\Pi_\perp^{{\rm(vac)}\mu\nu}$ be given by
\begin{eqnarray}
&&
\Pi_\perp^{{\rm(vac)}\mu\nu}(p_0,\vec{p})
=
g^{\mu\nu}
\left[ F_\pi^2(0) + p^2 \widetilde{\Pi}_\perp^S(p^2) \right]
+
(g^{\mu\nu} p^2 - p^\mu p^\nu )
\left[
  2 z_2(M_\rho) + \widetilde{\Pi}_\perp^{LT}(p^2)
\right]
\ ,
\label{PiA T0}
\end{eqnarray}
where $z_2(M_\rho)$ is renormalized at the scale $M_\rho$ and
$\widetilde{\Pi}_\perp^{LT}(p^2)$ is the finite renormalization
subject to the condition
\begin{equation}
\mbox{Re} \,\widetilde{\Pi}_\perp^{LT}(p^2=M_\rho^2) = 0 \ .
\end{equation}

To distinguish the hadronic thermal correction to the pion decay
constant from that to the parameter $z_2$, we decompose the
two-point function of $\overline{\cal A}_\mu$-$\overline{\cal
A}_\nu$ into four components as
\begin{equation}
 \Pi_\perp^{\mu\nu}=u^\mu u^\nu \Pi_\perp^t +
   (g^{\mu\nu}-u^\mu u^\nu)\Pi_\perp^s +
   P_L^{\mu\nu}\Pi_\perp^L + P_T^{\mu\nu}\Pi_\perp^T \ ,
\label{Pi perp decomp}
\end{equation}
where $P_L^{\mu\nu}$ and $P_T^{\mu\nu}$ are the polarization
tensors defined by
\begin{eqnarray}
  P_T^{\mu\nu}
&\equiv&
  g^\mu_i
  \left(
    \delta_{ij} - \frac{\vec{p}_i \vec{p}_j}{ \vert \vec{p} \vert^2}
  \right)
  g_j^\nu
\nonumber\\
&=&
  \left( g^{\mu\alpha} - u^\mu u^\alpha \right)
  \left(
    - g_{\alpha\beta} - \frac{p^\alpha p^\beta}{\bar{p}^2}
  \right)
  \left( g^{\beta\nu} - u^\beta u^\nu \right)
\ ,
\nonumber\\
  P_L^{\mu\nu}
&\equiv&
  - \left( g^{\mu\nu} - \frac{p^\mu p^\nu}{p^2} \right)
  - P_T^{\mu\nu}
\nonumber\\
&=&
  \left( g^{\mu\alpha} - \frac{p^\mu p^\alpha}{p^2} \right)
  u_\alpha
  \frac{p^2}{\vert\vec{p}\vert^2}
  u_\beta
  \left( g^{\beta\nu} - \frac{p^\beta p^\nu}{p^2} \right)
\ .
\label{pols}
\end{eqnarray}
Similarly to the division in Eq.~(\ref{Pi A div}),
it is convenient to divide each component into two parts as
\begin{equation}
\Pi_\perp^t(p_0,\bar{p};T) =
\Pi_\perp^{{\rm(vac)}t}(p_0,\bar{p}) +
\bar{\Pi}_\perp^t(p_0,\bar{p};T)
\ ,
\end{equation}
where $\Pi_\perp^{{\rm(vac)}t}(p_0,\vec{p})$ includes the tree
contribution plus the finite renormalization effect and
$\bar{\Pi}_\perp^t(p_0,\vec{p};T)$ is the hadronic thermal
contribution. {}With Eq.~(\ref{PiA T0}) the functions
$\Pi_\perp^{{\rm(vac)}t,s,L,T}$ can be written as
\begin{eqnarray}
&&
\Pi_\perp^{{\rm(vac)}t}(p_0,\bar{p}) =
\Pi_\perp^{{\rm(vac)}s}(p_0,\bar{p}) =
F_\pi^2(0) + \widetilde{\Pi}_\perp^S(p^2)
\ ,
\nonumber\\
&&
\Pi_\perp^{{\rm(vac)}L}(p_0,\bar{p}) =
\Pi_\perp^{{\rm(vac)}T}(p_0,\bar{p}) =
- p^2 \left[
  2 z_2(M_\rho) + \widetilde{\Pi}_\perp^{LT}(p^2)
\right]
\ .
\end{eqnarray}
The explicit forms of the hadronic thermal corrections
$\bar{\Pi}_\perp^{t,s,L,T}(p_0,\bar{p};T)$
are summarized in Eqs.~(\ref{AA t})--(\ref{AA T})
in Appendix~\ref{app:HTC}.

\begin{figure}
\begin{center}
\epsfxsize = 14cm
\ \epsfbox{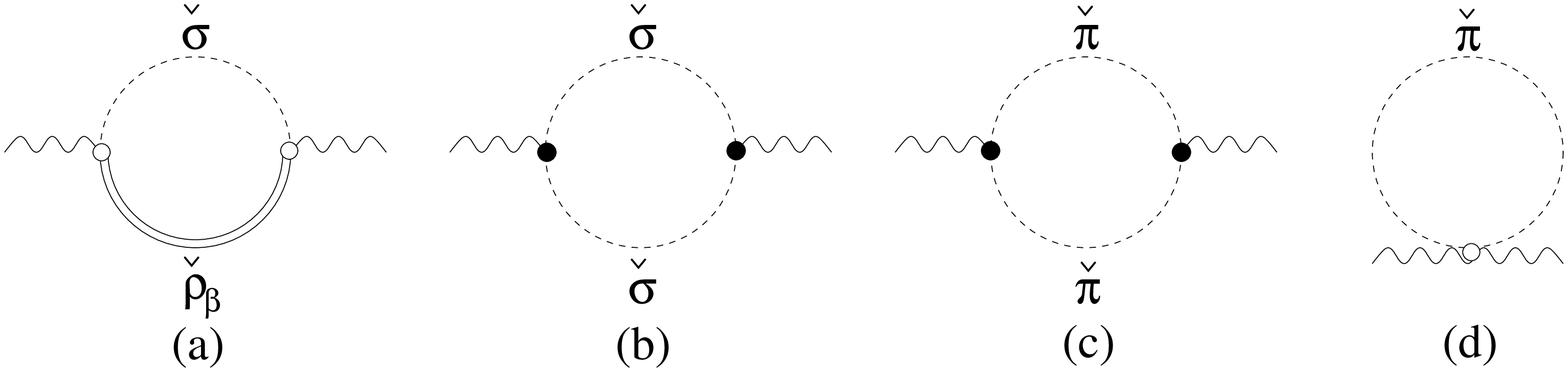}
\end{center}
\caption[]{%
Feynman diagrams contributing to the
$\overline{\cal V}_\mu$-$\overline{\cal V}_\nu$ two-point function.
Here $\check{\pi}$ represents the quantum pion field and likewise for
the others.}\label{fig:VV}
\end{figure}

We show the diagrams contributing to the
$\overline{\cal V}_\mu$-$\overline{\cal V}_\nu$ two-point function
in Fig.~\ref{fig:VV}.
We adopt
the on-shell renormalization condition similar to the above
to obtain
the resultant forms of four components of the two-point function
as
\begin{eqnarray}
&&
\Pi_{\parallel}^{{\rm(vac)}t}(p_0,\bar{p}) =
\Pi_{\parallel}^{{\rm(vac)}s}(p_0,\bar{p})
=
F_\sigma^2(M_\rho) + \widetilde{\Pi}_V^S(p^2)
\ ,
\nonumber\\
&&
\Pi_{\parallel}^{{\rm(vac)}L}(p_0,\bar{p}) =
\Pi_{\parallel}^{{\rm(vac)}T}(p_0,\bar{p})
=
-p^2 \left[
  2 z_1(M_\rho) + \widetilde{\Pi}_{\parallel}^{LT}(p^2)
\right]
\ ,
\label{PiVV vac}
\end{eqnarray}
where the parameters are renormalized at the scale $M_\rho$ and
the finite renormalization terms satisfy
\begin{eqnarray}
\mbox{Re} \,\widetilde{\Pi}_V^S(p^2=M_\rho^2)
=
\mbox{Re} \,\widetilde{\Pi}_{\parallel}^{LT}(p^2=M_\rho^2)
= 0
\ .
\label{cond V0}
\end{eqnarray}
For the two-point functions of
$\overline{V}_\mu$-$\overline{V}_\nu$ and
$\overline{V}_\mu$-$\overline{\cal V}_\nu$
we adopt similar on-shell
renormalization conditions. The resultant sums of the tree
contributions and quantum corrections take the forms
\begin{eqnarray}
&&
\Pi_V^{{\rm(vac)}t}(p_0,\bar{p}) =
\Pi_V^{{\rm(vac)}s}(p_0,\bar{p})
=
\Pi_{V\parallel}^{{\rm(vac)}t}(p_0,\bar{p}) =
\Pi_{V\parallel}^{{\rm(vac)}s}(p_0,\bar{p})
\nonumber\\
&& \quad
=
F_\sigma^2(M_\rho) + \widetilde{\Pi}_V^S(p^2)
=
\Pi_{\parallel}^{{\rm(vac)}t}(p_0,\bar{p}) =
\Pi_{\parallel}^{{\rm(vac)}s}(p_0,\bar{p})
\ ,
\nonumber\\
&&
\Pi_V^{{\rm(vac)}L}(p_0,\bar{p}) =
\Pi_V^{{\rm(vac)}T}(p_0,\bar{p})
=
-p^2 \left[
  - \frac{1}{g^2(M_\rho)} + \widetilde{\Pi}_V^{LT}(p^2)
\right]
\ ,
\nonumber\\
&&
\Pi_{V\parallel}^{{\rm(vac)}L}(p_0,\bar{p}) =
\Pi_{V\parallel}^{{\rm(vac)}T}(p_0,\bar{p})
=
-p^2 \left[
  z_3(M_\rho) + \widetilde{\Pi}_{V\parallel}^{LT}(p^2)
\right]
\ ,
\label{Pis vac}
\end{eqnarray}
where, as in Eq.~(\ref{cond V0}),
the finite renormalization terms satisfy
\begin{eqnarray}
\mbox{Re} \,\widetilde{\Pi}_V^S(p^2=M_\rho^2)
=
\mbox{Re} \,\widetilde{\Pi}_V^{LT}(p^2=M_\rho^2)
=
\mbox{Re} \,\widetilde{\Pi}_{V\parallel}^{LT}(p^2=M_\rho^2)
= 0
\ .
\label{cond V}
\end{eqnarray}
The hadronic thermal corrections to the above two-point functions
relevant to the present analysis are given in Eqs.~(\ref{rr vv rv
ts}) and (\ref{vv L}) in Appendix~\ref{app:HTC}.

It should be noticed that the renormalized parameters have the
intrinsic temperature dependences in addition to the dependence on the
renormalization point.
Then, the notations used above for the parameters renormalized at
on-shell should be understood as the following abbreviated notations:
\begin{eqnarray}
&& F_\pi(0) \equiv F_\pi(\mu=0;T) \ ,
\nonumber\\
&& F_\sigma(M_\rho) \equiv F_\sigma(\mu=M_\rho(T);T) \ ,
\nonumber\\
&& g(M_\rho) \equiv g(\mu=M_\rho(T);T) \ ,
\nonumber\\
&& z_{1,2,3}(M_\rho) \equiv z_{1,2,3}(\mu=M_\rho(T);T)
\ ,
\end{eqnarray}
where $\mu$ is the renormalization point and the mass parameter
$M_\rho$ is determined from the on-shell condition:
\begin{equation}
M_\rho^2 \equiv M_\rho^2(T)
= g^2(\mu=M_\rho(T);T) F_\sigma^2(\mu=M_\rho(T);T)
\ .
\end{equation}
In addition,
the parameter $a$ appearing in several expressions in Appendices
are defined as
\begin{equation}
a \equiv \frac{F_\sigma^2(\mu=M_\rho(T);T)}{F_\pi^2(\mu=M_\rho(T);T)}
\ .
\end{equation}

\section{Current Correlators}
\label{sec:CC}

\indent\indent
We now turn to construct the axial-vector and vector current
correlators from the two-point functions calculated in the previous
section.
The correlators are defined by
\begin{eqnarray}
G_A^{\mu\nu}(p_0=i\omega_n,\vec{p};T) \delta_{ab}
=
\int_0^{1/T} d \tau \int d^3\vec{x}
e^{-i(\vec{p}\cdot\vec{x}+\omega_n\tau)}
\left\langle
  J_{5a}^\mu(\tau,\vec{x}) J_{5b}^\nu(0,\vec{0})
\right\rangle_\beta
\ ,
\nonumber\\
G_V^{\mu\nu}(p_0=i\omega_n,\vec{p};T) \delta_{ab}
=
\int_0^{1/T} d \tau \int d^3\vec{x}
e^{-i(\vec{p}\cdot\vec{x}+\omega_n\tau)}
\left\langle
  J_a^\mu(\tau,\vec{x}) J_b^\nu(0,\vec{0})
\right\rangle_\beta
\ ,
\end{eqnarray}
where $J_{5a}^\mu$ and $J_a^\mu$ are, respectively, the
axial-vector and vector currents, $\omega_n=2n\pi T$ is the
Matsubara frequency, $(a,b)=1,\ldots,N_f^2-1$ denotes the flavor
index and $\langle ~\rangle_\beta$ the thermal average. The
correlators for Minkowski momentum are obtained by the analytic
continuation of $p_0$.

For constructing the axial-vector current correlator
$G_A^{\mu\nu}(p_0,\vec{p};T)$ from the
$\overline{\cal A}_\mu$-$\overline{\cal A}_\nu$ two-point function,
it is convenient to take the unitary gauge of the
background HLS and parameterize the background fields
$\bar{\xi}_{\rm L}$ and $\bar{\xi}_{\rm R}$ as
\begin{equation}
\bar{\xi}_{\rm L} = e^{-\bar{\phi}} \ ,
\quad
\bar{\xi}_{\rm R} = e^{\bar{\phi}} \ ,
\quad
\bar{\phi} = \bar{\phi}_a T_a \ .
\end{equation}
where $\bar{\phi}$ denotes the background field corresponding to
the pion field. In terms of this $\bar{\phi}$, the background
$\overline{\cal A}_\mu$ is expanded as
\begin{equation}
\overline{\cal A}_\mu
= {\cal A}_\mu + \partial_\mu \bar{\phi} + \cdots \ ,
\label{A bar exp}
\end{equation}
where the ellipses stand for the terms that include two or more
fields. Then, the axial-vector current correlator is
\begin{eqnarray}
G_A^{\mu\nu} =
\frac{
  p_\alpha p_\beta \Pi_\perp^{\mu\alpha} \Pi_\perp^{\nu\beta}
}{
  - p_{\bar{\mu}} p_{\bar{\nu}} \Pi_\perp^{\bar{\mu}\bar{\nu}}
}
+ \Pi_\perp^{\mu\nu}
\ ,
\end{eqnarray}
where the first term comes from the $\bar{\phi}$-exchange and the
second term from the direct ${\cal A}_\mu$-${\cal A}_\nu$
interaction. By using the decomposition in Eq.~(\ref{Pi perp
decomp}), this can be rewritten as
\begin{equation}
G_A^{\mu\nu} =
P_L^{\mu\nu} G_A^L + P_T^{\mu\nu} G_A^T
\ ,
\label{GA}
\end{equation}
where
\begin{eqnarray}
G_A^L
&=&
\frac{ p^2 \, \Pi_\perp^t \Pi_\perp^s }{
  - \left[
      p_0^2 \, \Pi_\perp^t - \bar{p}^2 \, \Pi_\perp^s
  \right]
}
+ \Pi_\perp^L
\ ,
\label{GAL}
\\
G_A^T
&=&
- \Pi_\perp^s + \Pi_\perp^T
\ .
\label{GAT}
\end{eqnarray}
One can see from Eqs.~(\ref{GAL}) and (\ref{GAT})
that the pion exchange contribution
is included only in the longitudinal component $G_A^L$.

To obtain the vector current correlator $G_V$, we first consider
the $\overline{V}$ propagator. By using the fact that the inverse
$\overline{V}$ propagator $i (D^{-1})^{\mu\nu}$ is equal to
$\Pi_V^{\mu\nu}$, the propagator for the field $\overline{V}$ can
be expressed as
\begin{eqnarray}
- i D_V^{\mu\nu}
=
u^\mu u^\nu D_V^t + ( g^{\mu\nu} - u^\mu u^\nu ) D_V^s
+ P_L^{\mu\nu} D_V^L + P_T^{\mu\nu} D_V^T
\ ,
\end{eqnarray}
where
\begin{eqnarray}
&&
D_V^t =
\frac{
  p^2 ( \Pi_V^s - \Pi_V^L )
}{
  p_0^2 \Pi_V^t (\Pi_V^s - \Pi_V^L)
  - \bar{p}^2 \Pi_V^s (\Pi_V^t - \Pi_V^L)
}
\ ,
\label{Dt def}
\\
&&
D_V^s =
\frac{
  p^2 ( \Pi_V^t - \Pi_V^L )
}{
  p_0^2 \Pi_V^t (\Pi_V^s - \Pi_V^L)
  - \bar{p}^2 \Pi_V^s (\Pi_V^t - \Pi_V^L)
}
\ ,
\label{Ds def}
\\
&&
D_V^L =
\frac{
  - p^2 \Pi_V^L
}{
  p_0^2 \Pi_V^t (\Pi_V^s - \Pi_V^L)
  - \bar{p}^2 \Pi_V^s (\Pi_V^t - \Pi_V^L)
}
\ ,
\label{DL def}
\\
&&
D_V^T =
D_V^s - \frac{1}{ \Pi_V^s - \Pi_V^T }
\ .
\label{DT def}
\end{eqnarray}
By using the above propagator $D_V$ and two-point functions of
$\overline{\cal V}_\mu$-$\overline{\cal V}_\mu$ and
$\overline{V}_\mu$-$\overline{\cal V}_\nu$, $G_V$ can be put into
the form
\begin{equation}
G_V^{\mu\nu} = \Pi_{V\parallel}^{\mu\alpha}
i D_{V,\alpha\beta} \Pi_{V\parallel}^{\beta\nu}
+ \Pi_{\parallel}^{\mu\nu}
\ .
\end{equation}
After a lengthy calculation, we obtain
\begin{eqnarray}
G_V^{\mu\nu}
&=&
u^\mu u^\nu
\Biggl[
  \frac{ D_V^L }{ \Pi_V^L }
  \biggl\{
    \frac{\bar{p}^2}{ p^2 } \Pi_{V\parallel}^L
    \left(
      \Pi_V^s \Pi_{V\parallel}^t - \Pi_V^t \Pi_{V\parallel}^s
    \right)
\nonumber\\
&& \qquad\qquad
    {} - \frac{\Pi_{V\parallel}^t}{p^2}
    \biggl(
      -p_0^2 \Pi_{V\parallel}^t (\Pi_V^s - \Pi_V^L)
      + \bar{p}^2 \left(
        \Pi_{V\parallel}^t \Pi_V^s - \Pi_{V\parallel}^s \Pi_V^L
      \right)
    \biggr)
  \biggr\}
  + \Pi_\parallel^t
\Biggr]
\nonumber\\
&&
{} + ( g^{\mu\nu} - u^\mu u^\nu )
\Biggl[
  \frac{ D_V^L }{ \Pi_V^L }
  \biggl\{
    \frac{p_0^2}{ p^2 } \Pi_{V\parallel}^L
    \left(
      \Pi_V^s \Pi_{V\parallel}^t - \Pi_V^t \Pi_{V\parallel}^s
    \right)
\nonumber\\
&& \qquad\qquad
    {} - \frac{\Pi_{V\parallel}^s}{p^2}
    \biggl(
      -p_0^2 \left(
        \Pi_V^t \Pi_{V\parallel}^s - \Pi_{V\parallel}^t\Pi_V^L
      \right)
      + \bar{p}^2 \Pi_{V\parallel}^s \left(
        \Pi_V^t - \Pi_V^L
      \right)
    \biggr)
  \biggr\}
  + \Pi_\parallel^s
\Biggr]
\nonumber\\
&&
{} + P_L^{\mu\nu}
\Biggl[
  \frac{ D_V^L }{ \Pi_V^L }
  \biggl\{
    - \Pi_V^L \Pi_{V\parallel}^t \Pi_{V\parallel}^s
    + \Pi_{V\parallel}^L \left(
      \Pi_{V\parallel}^t \Pi_V^s + \Pi_{V\parallel}^s \Pi_V^t
    \right)
\nonumber\\
&& \qquad\qquad
    {} - \frac{1}{p^2} \left(
      p_0^2 \Pi_V^t - \bar{p}^2 \Pi_V^s
    \right)
    \left( \Pi_{V\parallel}^L \right)^2
  \biggr\}
  + \Pi_\parallel^L
\Biggr]
\nonumber\\
&&
{} + P_T^{\mu\nu}
\Biggl[
  \frac{ D_V^L }{ \Pi_V^L }
  \biggl\{
    - \frac{p_0^2}{p^2} \Pi_{V\parallel}^L
    \left(
      \Pi_V^t \Pi_{V\parallel}^s - \Pi_V^s \Pi_{V\parallel}^t
    \right)
\nonumber\\
&& \qquad\qquad
    {} - \frac{\Pi_{V\parallel}^s}{p^2}
    \left(
      -p_0^2 \left(
        \Pi_V^t \Pi_{V\parallel}^s - \Pi_{V\parallel}^t\Pi_V^L
      \right)
      + \bar{p}^2 \Pi_{V\parallel}^s \left(
        \Pi_V^t - \Pi_V^L
      \right)
    \right)
  \biggr\}
\nonumber\\
&& \qquad\qquad
  {} + \frac{ \left( \Pi_{V\parallel}^s - \Pi_{V\parallel}^T \right)^2
  }{ \Pi_V^s - \Pi_V^T }
  + \Pi_\parallel^T
\Biggr]
\ .
\label{PGP exp 1}
\end{eqnarray}
One might worry that the above form does not satisfy the current
conservation $p_\mu G_V^{\mu\nu} = 0$. However since, as shown in
Eq.~(\ref{rr vv rv ts}), the conditions
\begin{eqnarray}
&& \Pi_V^t = - \Pi_{V\parallel}^t = \Pi_{\parallel}^t \ ,
\nonumber\\
&&
\Pi_V^s = - \Pi_{V\parallel}^s = \Pi_{\parallel}^s
\label{t s equality}
\end{eqnarray}
are satisfied, Eq.~(\ref{PGP exp 1}) can be rewritten as
\begin{eqnarray}
G_V^{\mu\nu}
&=&
P_L^{\mu\nu}
\left[
  \left( \frac{ - D_V^L }{ \Pi_V^L } \right)
  \left\{
    \Pi_V^t \Pi_V^s \left( \Pi_V^L + 2 \Pi_{V\parallel}^L \right)
    + \frac{ p_0^2 \Pi_V^t - \bar{p}^2 \Pi_V^s }{p^2}
    \left( \Pi_{V\parallel}^L \right)^2
  \right\}
  + \Pi_{\parallel}^L
\right]
\nonumber\\
&&
{} +
P_T^{\mu\nu}
\left[
  \frac{
    \Pi_V^s \left( \Pi_V^T + 2 \Pi_{V\parallel}^T \right)
    + \left( \Pi_{V\parallel}^T \right)^2
  }{
    \Pi_V^s - \Pi_V^T
  }
  + \Pi_{\parallel}^T
\right]
\ .
\label{GV mn form}
\end{eqnarray}
Now it is evident that the current is conserved since $p_\mu
P_L^{\mu\nu} = p_\mu P_T^{\mu\nu} = 0$. In the present analysis,
the equality $\bar{\Pi}_V^t = \bar{\Pi}_V^s$ is seen to hold as
shown in Eq.~(\ref{rr vv rv ts}). This implies that $\Pi_V^t =
\Pi_V^s$ is also satisfied since the quantum corrections to
$\Pi_V^t$ and $\Pi_V^s$ are equal to each other due to Lorentz
invariance. Thus, $G_V^{\mu\nu}$ can be written as
\begin{equation}
G_V^{\mu\nu} = P_L^{\mu\nu} G_V^L + P_T^{\mu\nu} G_V^T
\ ,
\label{GV mn}
\end{equation}
where
\begin{eqnarray}
G_V^L
&=&
  \frac{
    \Pi_V^t \left( \Pi_V^L + 2 \Pi_{V\parallel}^L \right)
  }{
    \Pi_V^t - \Pi_V^L
  }
  + \Pi_{\parallel}^L
\label{GVL}
\\
G_V^T
&=&
  \frac{
    \Pi_V^t \left( \Pi_V^T + 2 \Pi_{V\parallel}^T \right)
  }{
    \Pi_V^t - \Pi_V^T
  }
  + \Pi_{\parallel}^T
\ .
\label{GVT}
\end{eqnarray}
Note that, in the above expressions, we have dropped the terms
$\left( \Pi_{V\parallel}^L \right)^2$ and
$\left( \Pi_{V\parallel}^T \right)^2$ since they are of higher order.

\section{Pion Decay Constants and Pion Velocity}
\label{sec:PDCV}
 \indent\indent We now proceed to study the
on-shell structure of the pion. For this we look at the pole of
the longitudinal component $G_A^L$ in Eq.~(\ref{GAL}). Since both
$\Pi_\perp^t$ and $\Pi_\perp^s$ have imaginary parts, we choose to
determine the pion energy $E$ from the real part by solving the
dispersion formula
\begin{eqnarray}
  0
&=&
  \left[
    p_0^2 \, \mbox{Re} \Pi^{t}_\perp (p_0,\bar{p};T)
    - \bar{p}^2 \, \mbox{Re} \Pi^{s}_\perp (p_0,\bar{p};T)
  \right]_{p_0=E}
\ ,
\label{pi on shell cond}
\end{eqnarray}
where $\bar{p}\equiv\vert\vec{p}\vert$. As remarked in
Section~\ref{sec:TPF}, in HLS at one-loop level, $\Pi^{t}_\perp
(p_0,\bar{p};T)$ and $\Pi^{s}_\perp (p_0,\bar{p};T)$ are of the
form
\begin{eqnarray}
\Pi^{t}_\perp (p_0,\bar{p};T)
&=&
F_\pi^2(0) + \widetilde{\Pi}_\perp^S(p^2) +
\bar{\Pi}^{t}_\perp (p_0,\bar{p};T)
\ ,
\nonumber\\
\Pi^{s}_\perp (p_0,\bar{p};T)
&=&
F_\pi^2(0) + \widetilde{\Pi}_\perp^S(p^2) +
\bar{\Pi}^{s}_\perp (p_0,\bar{p};T)
\ ,
\label{Pi t s forms}
\end{eqnarray}
where $\widetilde{\Pi}_\perp^S(p^2)$ is the finite renormalization
contribution, and $\bar{\Pi}^{t}_\perp (p_0,\bar{p};T) $ and
$\bar{\Pi}^{s}_\perp (p_0,\bar{p};T) $ are the hadronic thermal
contributions. Substituting Eq.~(\ref{Pi t s forms}) into
Eq.~(\ref{pi on shell cond}), we obtain
\begin{equation}
0 =
\left( E^2 - \bar{p}^2 \right)
\left[
  F_\pi^2(0) +
  \mbox{Re} \,\widetilde{\Pi}_\perp^S( p^2=E^2 - \bar{p}^2 )
\right]
+
  E^2
  \mbox{Re} \,\bar{\Pi}^{t}_\perp (E,\bar{p};T)
  - \bar{p}^2
  \mbox{Re} \, \bar{\Pi}^{s}_\perp (E,\bar{p};T)
\ .
\end{equation}
The pion velocity $v_\pi(\bar{p}) \equiv E / \bar{p}$ is then obtained
by solving
\begin{eqnarray}
v_\pi^2(\bar{p})
&=&
\frac{
  F_\pi^2(0) +
  \mbox{Re} \, \bar{\Pi}^{s}_\perp (\bar{p},\bar{p};T)
}{
  F_\pi^2(0) +
  \mbox{Re} \, \bar{\Pi}^{t}_\perp (\bar{p},\bar{p};T)
}
\ .
\label{v2: form}
\end{eqnarray}
Here we replaced $E$ by $\bar{p}$ in the hadronic thermal terms
$\bar{\Pi}_\perp^t (E,\vec{p})$ and $\bar{\Pi}_\perp^s
(E,\vec{p})$ as well as in the finite renormalization contribution
$\widetilde{\Pi}_\perp^S( p^2 = E^2 - \bar{p}^2)$, since the
difference is of higher order. [Note that $
\widetilde{\Pi}_\perp^S( p^2=0 )= 0$.]

Next we determine the wave function renormalization of the pion
field, which relates the background field $\bar{\phi}$ to the pion
field $\bar{\pi}$ in the momentum space as
 \be
\bar{\phi} = \bar{\pi}/\widetilde{F}(\bar{p};T).
 \ee
We follow the analysis in Ref.~\cite{MOW} to obtain
\begin{equation}
\widetilde{F}^2(\bar{p};T) =
\mbox{Re} \Pi_\perp^t(E,\bar{p};T)
= F_\pi^2(0) +
  \mbox{Re} \, \bar{\Pi}^{t}_\perp (\bar{p},\bar{p};T)
\ .
\label{Ftil def}
\end{equation}
Using this wave function renormalization and the velocity in
Eq.~(\ref{v2: form}), we can rewrite the longitudinal part of the
axial-vector current correlator as
\begin{eqnarray}
G_A^L(p_0,\vec{p})
&=&
\frac{ p^2
  \Pi_\perp^t(p_0,\bar{p};T) \Pi_\perp^s(p_0,\bar{p};T)
  / \widetilde{F}^2(\bar{p};T)
}{
  - \left[
      p_0^2 - v_\pi^2(\bar{p}) \bar{p}^2 + \Pi_\pi(p_0,\bar{p};T)
  \right]
}
+ \Pi_\perp^L(p_0,\bar{p};T)
\ ,
\label{GAL 2}
\end{eqnarray}
where the pion self energy $\Pi_\pi(p_0,\bar{p};T)$ is given by
\begin{eqnarray}
&&
\Pi_\pi(p_0,\bar{p};T)
=
\frac{1}{
  \mbox{Re} \, \Pi^{t}_\perp (E,\bar{p};T)
}
\nonumber\\
&& \quad
\times
\Biggl[
  p_0^2
  \left\{
    \Pi^{t}_\perp (p_0,\bar{p};T)
    -
    \mbox{Re} \, \Pi^{t}_\perp (E,\bar{p};T)
  \right\}
  -
  \bar{p}^2
  \left\{
    \Pi^{s}_\perp (p_0,\bar{p};T)
    -
    \mbox{Re} \, \Pi^{s}_\perp (E,\bar{p};T)
  \right\}
\Biggr]
\ .
\end{eqnarray}

Let us now define the pion decay constant. A natural procedure is
to define the pion decay constant from the pole residue of the
axial-vector current correlator. From Eq.~(\ref{GAL 2}), the pion
decay constant is given by
\begin{eqnarray}
f_\pi^2(\bar{p};T)
&=&
\frac{
 \Pi_\perp^t(E,\bar{p};T)
 \Pi_\perp^s(E,\bar{p};T)
}{ \widetilde{F}^2(\bar{p};T) }
\nonumber\\
&=&
\frac{
 \left[ F_\pi^2(0) + \bar{\Pi}^{t}_\perp (\bar{p},\bar{p};T) \right]
 \left[ F_\pi^2(0) + \bar{\Pi}^{s}_\perp (\bar{p},\bar{p};T) \right]
}{ \widetilde{F}^2(\bar{p};T) }
\ .
\label{fpi2 def}
\end{eqnarray}
We now address how $f_\pi^2(\bar{p};T)$ is related to the temporal
and spatial components of the pion decay constant introduced in
Ref.~\cite{PT:96}. Following their notation, let $f_\pi^t$ denote
the decay constant associated with the temporal component of the
axial-vector current and $f_\pi^s$ the one with the spatial
component. In the present analysis, they can be read off from the
coupling of the $\bar{\pi}$ field to the axial-vector external
field ${\cal A}_\mu$:
\begin{eqnarray}
  f_\pi^t(\bar{p};T)
&\equiv&
  \frac{
    \Pi^{t}_\perp (E,\bar{p};T)
  }{\widetilde{F}(\bar{p};T)}
=
  \frac{ F_\pi^2(0) + \bar{\Pi}^{t}_\perp (\bar{p},\bar{p};T) }
    {\widetilde{F}(\bar{p};T)}
\ ,
\label{fpit def}
\\
  f_\pi^s(\bar{p};T)
&\equiv&
  \frac{
    \Pi^{s}_\perp (\tilde{E},\bar{p};T)
  }{\widetilde{F}(\bar{p};T)}
=
  \frac{ F_\pi^2(0) + \bar{\Pi}^{s}_\perp (\bar{p},\bar{p};T) }
    {\widetilde{F}(\bar{p};T)}
\ .
\label{fpis def}
\end{eqnarray}
Comparing Eqs.~(\ref{fpit def}) and (\ref{fpis def}) with
Eqs.~(\ref{v2: form}), (\ref{Ftil def}) and (\ref{fpi2 def}), we
have~\cite{PT:96,MOW}
\begin{eqnarray}
\widetilde{F}(\bar{p};T)
&=& \mbox{Re} \, f_\pi^t(\bar{p};T)
\ ,
\\
f_\pi^2(\bar{p};T) &=&
  f_\pi^t(\bar{p};T) f_\pi^s(\bar{p};T)
\ ,
\\
v_\pi^2(\bar{p}) &=& \frac{\mbox{Re}
\,f_\pi^s(\bar{p};T)}{\mbox{Re}\,f_\pi^t(\bar{p};T)} \ .
\label{v2 rel}
\end{eqnarray}

We are now ready to investigate what happens to the above
quantities when the critical temperature $T_c$ is approached. Due
to the VM in hot matter~\cite{HaradaSasaki},  the $parametric$
$\rho$ meson mass goes to zero ($M_\rho\rightarrow0$) and the
parameter $a$ approaches one ($a\rightarrow1$), so we have [see
Eq.~(\ref{PiA ts Tc app})]
\begin{eqnarray}
&&
\bar{\Pi}_\perp^t(\bar{p},\bar{p};T)
\mathop{\longrightarrow}_{T \rightarrow T_c}
- \frac{N_f}{2} \widetilde{J}_{1}^2(0;T_c)
= - \frac{N_f}{24} T_c^2
\ ,
\nonumber\\
&&
\bar{\Pi}_\perp^s(\bar{p},\bar{p};T)
\mathop{\longrightarrow}_{T \rightarrow T_c}
- \frac{N_f}{2} \widetilde{J}_{1}^2(0;T_c)
= - \frac{N_f}{24} T_c^2
\ .
\label{PiA ts Tc}
\end{eqnarray}
Substituting these into the expression of the pion velocity in
Eq.~(\ref{v2: form}), we obtain
\begin{equation}
v_\pi^2(\bar{p}) \mathop{\longrightarrow}_{T \rightarrow T_c} 1 \
.
\end{equation}
This is our first main result: in the framework of the VM, the
pion velocity approaches 1 near the critical
temperature~\footnote{Modulo small Lorentz-breaking correction
mentioned above.}, not 0 as in the case of the pion-only
situation~\cite{SS:1,SS:2}.

{}From Eq.~(\ref{PiA ts Tc}), we can evaluate the pion decay
constant Eq.~(\ref{fpi2 def}) at the critical temperature which
comes out to be
\begin{equation}
f_\pi^2(\bar{p};T_c) = F_\pi^2(0) - \frac{N_f}{24} T_c^2
\ .
\end{equation}
Since this $f_\pi$ is the order parameter and should vanish at the
critical temperature, the parameter $F_\pi^2(0)$ at $T=T_c$ is
given at $T_c$ as~\cite{HaradaSasaki}
\begin{equation}
F_\pi^2(0)
\ \mathop{\longrightarrow}_{T \rightarrow T_c}\
\frac{N_f}{24} T_c^2
\ .
\label{Fp0 Tc}
\end{equation}
Substituting Eq.~(\ref{PiA ts Tc}) together with Eq.~(\ref{Fp0
Tc}) into Eqs.~(\ref{fpit def}) and (\ref{fpis def}), we conclude
that both temporal and spatial pion decay constants vanish at the
critical temperature~\footnote{These results differ from those
obtained in a framework in which the {\it only} relevant degrees of
freedom near chiral restoration are taken to be the
pions~\cite{SS:1,SS:2}. We will explain how this comes about in
the conclusion section.}:
\begin{equation}
f_\pi^t(\bar{p};T_c) = f_\pi^s(\bar{p};T_c) = 0 \ .
\end{equation}
This is our second main result. Note that while $f_\pi^t (T_c)=0$,
$\chi_A (T_c)$ is non-zero in consistency with the lattice result.
Here the HLS gauge boson plays an essential role.

\section{Axial-Vector and Vector Susceptibilities}
\label{sec:SUS}
 \indent\indent In terms of the quantities defined
in the preceding sections, the axial-vector susceptibility
$\chi_A(T)$ and the vector susceptibility $\chi_V(T)$ for
non-singlet currents~\footnote{We will confine ourselves to
non-singlet (that is, isovector) susceptibilities, so we won't
specify the isospin structure from here on.} are given by the
$00$-component of the axial-vector and vector current correlators
in the static--low-momentum limit:
\begin{eqnarray}
&&
\chi_A(T)  = 2 N_f \,
  \lim_{\bar{p}\rightarrow0}
  \lim_{p_0\rightarrow0}
  \left[ G_A^{00}(p_0,\vec{p};T) \right]
\ ,
\nonumber\\
&&
\chi_V(T)
= 2 N_f \, \lim_{\bar{p}\rightarrow0}
  \lim_{p_0\rightarrow0}
  \left[ G_V^{00}(p_0,\vec{p};T) \right]
\ ,
\label{def chiV}
\end{eqnarray}
where we have included the normalization factor of $2 N_f$. Using
the current correlators given in Eqs.~(\ref{GA}) and (\ref{GV mn
form}) and noting that $ \lim_{p_0\rightarrow0} P_L^{00} =
\lim_{p_0\rightarrow0} \bar{p}^2/p^2 = - 1 $, we can express
$\chi_A(T)$ and $\chi_V(T)$ as
\begin{eqnarray}
&&
\chi_A(T)  = - 2 N_f \,\lim_{\bar{p}\rightarrow0}
\lim_{p_0\rightarrow0}
\left[
  \Pi_\perp^L(p_0,\vec{p};T) - \Pi_\perp^t(p_0,\vec{p};T)
\right]
\ ,
\nonumber\\
&&
\chi_V(T)  = - 2 N_f \,\lim_{\bar{p}\rightarrow0}
\lim_{p_0\rightarrow0}
\left[
  \frac{
    \Pi_V^t \left( \Pi_V^L + 2 \Pi_{V\parallel}^L \right)
  }{
    \Pi_V^t - \Pi_V^L
  }
  + \Pi_{\parallel}^L
\right]
\ ,
\end{eqnarray}
where for simplicity of notation, we have suppressed the argument
$(p_0,\vec{p};T)$ in the right-hand-side of the expression for
$\chi_V(T)$. In HLS theory at one-loop level, the susceptibilities
read
\begin{eqnarray}
&&
\chi_A(T)  = 2 N_f \left[
  F_\pi^2(0)
  + \lim_{\bar{p}\rightarrow0}
  \lim_{p_0\rightarrow0}
  \left\{
    \bar{\Pi}_\perp^t(p_0,\vec{p};T) -
    \bar{\Pi}_\perp^L(p_0,\vec{p};T)
  \right\}
\right]
\ ,
\nonumber\\
&&
\chi_V(T)  =
- 2 N_f \,\lim_{\bar{p}\rightarrow0}
\lim_{p_0\rightarrow0}
\left[
  \frac{
    \left( a(0) F_\pi^2(0) + \bar{\Pi}_V^t \right)
    \left( \bar{\Pi}_V^L + 2 \bar{\Pi}_{V\parallel}^L \right)
  }{
    a(0) F_\pi^2(0) + \bar{\Pi}_V^t
    - \bar{\Pi}_V^L
  }
  + \Pi_{\parallel}^L
\right]
\ ,
\label{chiV}
\end{eqnarray}
where the parameter $a(0)$ is defined by
\begin{eqnarray}
a(0) &=&
\frac{ \Pi_V^{{\rm(vac)}t}(p_0=0,\bar{p}=0) }{ F_\pi^2(0) }
= \frac{ \Pi_V^{{\rm(vac)}s}(p_0=0,\bar{p}=0) }{ F_\pi^2(0) }
\ .
\end{eqnarray}
In Ref.~\cite{HY:WM}, $a(0)$ was defined by the ratio
$F_\sigma^2(M_\rho)/F_\pi^2(0)$ without taking into account the
finite renormalization effect which depends on the details of the
renormalization condition. In the present renormalization
condition (\ref{Pis vac}) with Eq.~(\ref{cond V}), the finite
renormalization effect leads to
\begin{equation}
\widetilde{\Pi}_V^S(p^2=0) =
\frac{N_f}{(4\pi)^2} M_\rho^2
\left( 2 - \sqrt{3} \tan^{-1} \sqrt{3} \right)
\ ,
\end{equation}
and then $a(0)$ reads
\begin{equation}
a(0) = \frac{F_\sigma^2(M_\rho)}{F_\pi^2(0)}
+ \frac{N_f}{(4\pi)^2} \frac{ M_\rho^2 }{F_\pi^2(0)}
\left( 2 - \sqrt{3} \tan^{-1} \sqrt{3} \right)
\ .
\label{a0 exp}
\end{equation}
It follows from the static--low-momentum limit of
$(\bar{\Pi}_\perp^t - \bar{\Pi}_\perp^L)$ given in Eq.~(\ref{PiA
tmL SL}) that the axial-vector susceptibility $\chi_A(T)$ takes
the form
\begin{eqnarray}
\chi_A(T) &=& 2N_f
\Biggl[
  F_\pi^2(0)
  - N_f \widetilde{J}_{1}^2(0;T)
  + N_f a\, \widetilde{J}_{1}^2(M_\rho;T)
\nonumber\\
&& \qquad
  {}- N_f \frac{a}{M_\rho^2}
  \left\{
    \widetilde{J}_{-1}^2(M_\rho;T)
    - \widetilde{J}_{-1}^2(0;T)
  \right\}
\Biggr]
\ .
\label{chiA}
\end{eqnarray}
Near the critical temperature ($T\rightarrow T_c$), we have
$M_\rho \rightarrow 0$, $a\rightarrow1$ due to the intrinsic
temperature dependence in the VM in hot
matter~\cite{HaradaSasaki}.
Furthermore, from Eq.~(\ref{Fp0 Tc}), we see that the
parameter $F_\pi^2(0)$ approaches
$\frac{N_f}{24} T_c^2$ for $T\rightarrow T_c$. Substituting these
conditions into Eq.~(\ref{chiA}) and noting that
\begin{equation}
\lim_{M_\rho\rightarrow0}
\left[
  {}- \frac{1}{M_\rho^2}
  \left\{
    \widetilde{J}_{-1}^2(M_\rho;T)
    - \widetilde{J}_{-1}^2(0;T)
  \right\}
\right]
= \frac{1}{2} \widetilde{J}_{1}^2(0;T) = \frac{1}{24} T^2 \ ,
\end{equation}
we obtain
\begin{equation}
\chi_A(T_c) = \frac{N_f^2}{6} T_c^2
\ .
\label{axial SUS}
\end{equation}

To obtain the vector susceptibility near the critical temperature,
we first consider $ a(0) F_\pi^2(0) + \bar{\Pi}_V^t$ appearing in the
numerator of the first term in the right-hand-side of
Eq.~(\ref{chiV}). Using Eq.~(\ref{Pir t SL}), we get for the
static--low-momentum limit of $ a(0) F_\pi^2(0) + \bar{\Pi}_V^t$ as
\begin{eqnarray}
&&
\lim_{\bar{p}\rightarrow0}
\lim_{p_0\rightarrow0}
\left[ a(0) F_\pi^2(0) + \bar{\Pi}_V^t(p_0,\bar{p};T) \right]
\nonumber\\
&& \qquad
=
a(0) F_\pi^2(0)
- \frac{N_f}{4} \left[
  2 \widetilde{J}_{-1}^0(M_\rho;T)
  - \widetilde{J}_{1}^2(M_\rho;T)
  + a^2 \, \widetilde{J}_{1}^2(0;T)
\right]
\ .
\label{aF Pit SL}
\end{eqnarray}
{}From Eq.~(\ref{a0 exp}) we can see that $a(0) \rightarrow 1$ as
$T\rightarrow T_c$ since $F_\sigma^2(M_\rho) \rightarrow
F_\pi^2(0)$ and $M_\rho \rightarrow 0$. Furthermore,
$F_\pi^2(0)\rightarrow \frac{N_f}{24}T_c^2$ as we have shown in
Eq.~(\ref{Fp0 Tc}). Then, the first term of Eq.~(\ref{aF Pit SL})
approaches $\frac{N_f}{24}T_c^2$. The second term, on the other
hand, approaches $- \frac{N_f}{24}T_c^2$ as $M_\rho \rightarrow0$
and $a \rightarrow1$ for $T\rightarrow T_c$. Thus, we have
\begin{equation}
\lim_{\bar{p}\rightarrow0}
\lim_{p_0\rightarrow0}
\left[ a(0) F_\pi^2(0) + \bar{\Pi}_V^t(p_0,\bar{p};T) \right]
\mathop{\longrightarrow}_{T \rightarrow T_c}
0 \ .
\label{aF Pit 0}
\end{equation}
This implies that
only the second term $\Pi_\parallel^L$
in the right-hand-side of Eq.~(\ref{chiV}) contributes
to the vector susceptibility
near the critical temperature.
Thus, taking $M_\rho \rightarrow 0$ and $a\rightarrow1$,
in Eq.~(\ref{Piv L SL}),
we obtain
\begin{equation}
\chi_V(T_c) = \frac{N_f^2}{6} T_c^2
\ ,
\label{vector SUS}
\end{equation}
which agrees with the axial-vector susceptibility in
Eq.~(\ref{axial SUS}). This is a prediction, not an input
condition, of the theory. For $N_f=2$, we have
\begin{equation}
\chi_A(T_c) =
\chi_V(T_c) = \frac{2}{3} T_c^2
\ .
\label{SUS value}
\end{equation}
This is our third main result. The result $\chi_V(T_c) =
\frac{2}{3} T_c^2$ is consistent with the lattice result as
interpreted in \cite{Brown-Rho:96}. It is interesting to note that
the RPA result obtained in \cite{Brown-Rho:96} in NJL model in
terms of a quasi-quark-quasi-antiquark bubble is reproduced
quantitatively by the one-loop graphs in HLS with the VM.

It should be noticed that the pion pole effect does not
contribute to the ASUS in Eq.~(\ref{axial SUS})
since the pion decay constant
$f_\pi^t$ vanishes at the critical temperature as we have shown in the
previous section, and that
the contribution to the ASUS
comes from the
non-pole contribution expressed in Fig.~\ref{fig:AA}.
In three diagrams, the third diagram in Fig.~\ref{fig:AA}(c)
is proportional to $(1-a)$ and then it vanishes at the
critical temperature due to the VM.
Similarly, since the transverse $\rho$ decouples
at the critical point in the VM~\cite{HY:VM,HaradaSasaki},
the first diagram in Fig.~\ref{fig:AA}(a) does not contribute.
Thus,
the above result for the ASUS in Eq.~(\ref{axial SUS})
comes from only the contribution
generated via the (longitudinal) vector
meson plus pion loop [see Fig.~\ref{fig:AA}(b)].~\footnote{%
  Note that $\check{\sigma}$ in Fig.~\ref{fig:AA}(b) is the
  quantum field corresponding to the NG boson absorbed into
  the vector meson, i.e., the longitudinal vector meson.
}
Similarly,
the vector meson pole effect to the VSUS
vanishes at the critical
temperature as shown in Eq.~(\ref{aF Pit 0}),
and the contribution
to the VSUS is generated via the pion loop
[Fig.~\ref{fig:VV}(c)]
and
the longitudinal vector meson loop
[Fig.~\ref{fig:VV}(b)].~\footnote{%
  Note that the contribution from Fig.~\ref{fig:VV}(a) vanishes since
  the transverse $\rho$ decouples and that the one from
  Fig.~\ref{fig:VV}(d) also vanishes since it is proportional
  to $(1-a)$.
} Since the longitudinal vector meson becomes massless, degenerate
with the pion as the chiral partner in the VM, loop contribution
to the ASUS becomes identical to that to the VSUS. Thus, the
massless vector meson predicted by the VM fixed point plays an
essential role to obtain the above equality between the ASUS and
the VSUS.

In the present analysis, our aim is
to show the qualitative structure of the ASUS and the VSUS
in the VM,
i.e., {\it the equality between them is predicted by the VM}.
In order to compare our qualitative results with the lattice result,
we need to
go beyond the one-loop approximation.
We note here that there is a
result from a hard thermal loop calculation which gives $\chi_V
(T_c)\approx 1.3 T_c^2$~\cite{CMT}. However this result cannot be
compared to ours for two reasons. First we need to sum higher
loops in our formalism which may be done in random phase
approximation as in \cite{Kunihiro:91}. Second, the perturbative
QCD with a hard thermal loop approximation may not be valid in the
temperature regime we are considering. Even at $T\gg T_c$, the
situation is not clear as pointed out in Ref.~\cite{blaizot}.

\section{Summary and Remarks}
\label{sec:SR} \indent\indent The notion of the vector
manifestation in chiral symmetry requires that the zero-mass
vector mesons be present at the chiral phase transition. As
discussed by Brown and Rho~\cite{QM2002}, the light vector mesons
near the transition point ``bottom-up" can be considered as
Higgsed gluons in the sense of color-flavor locking in the broken
chiral symmetry sector proposed by Berges and
Wetterich~\cite{wett,berges-wett} and could figure in heavy-ion
processes measured at RHIC energies. In this paper we are finding
that in the VM, the vector mesons with vanishing masses at the
chiral transition (in the chiral limit) can figure importantly in
the vector and axial-vector susceptibilities near the chiral
transition point. The notable results are that the VM confirms
explicitly the equality $\chi_V=\chi_A$ at $T_c$ and that both
$f_\pi^t$ and $f_\pi^s$ vanish simultaneously with the pion
velocity $v_\pi\sim 1$. These differ from the results expected in
a scenario where only the pions are the relevant effective degrees
of freedom.

The reason for that the $v_\pi$ deviates from 1 is due to the
Lorentz-breaking term in the bare HLS Lagrangian at which the
matching to QCD is made in a thermal bath. We find the deviation
is small (this will be detailed in \cite{HKRS:prep}). The
small deviation from 1 is also found in dense skyrmion matter
studied in \cite{LPRV}.

If one assumes that the only light degrees of freedom near $T_c$
are the pions,
then one can simply take the current algebra terms in
the Lagrangian and the axial-vector susceptibility (ASUS) $\chi_A$
is uniquely given by the temporal component of the pion decay
constant $f_\pi^t$ with the degrees of freedom that are integrated
out renormalizing this constant. Then the unquestionable equality
$\chi_V=\chi_A$ at $T_c$ together with the lattice result
$\chi_V|_{T_C}\neq 0$ leads to the Son-Stephanov result on the
pion velocity $v_\pi=0$. There is however a caveat to this simple
result and it is that the same reasoning fails when one computes
explicitly the vector susceptibility (VSUS) using the same current
algebra Lagrangian.

Positing that the vector mesons enter in the VM near $T_c$
circumvents this caveat and at the same time, makes a concrete
prediction. In this framework, the ASUS is given by a term related
to $f_\pi^t$ plus contributions from the vector-meson (i.e., the
longitudinal component $\sigma$) loop. At $T_c$, the $f_\pi^t$
vanishes and what remains comes out precisely equal to the VSUS
$\chi_V$ in which the $\sigma$ (longitudinal $\rho$) loop in
$\chi_A$ is replaced by the pion loop. All these are perfectly
understandable in terms of the VM in HLS.

If chiral symmetry restoration \`a la HLS/VM -- but not the
standard chiral theory one -- is the valid scenario as is argued in
\cite{HY:PR}, its confirmation would provide a valuable insight
into some of the basic tenets of effective field theories as
expounded in \cite{HY:PR}.

As was shown in, e.g., Ref.~\cite{Brown-Rho:91,Brown-Rho:96}, we
expect that baryons become light very near the phase transition
point. At least when matter density is involved, the light baryons
must figure importantly. In \cite{HKR}, we have simulated density
effects by introducing quasiquarks whose effective mass is
expected to lie below the in-medium vector meson mass $m_\rho^*$.
In \cite{Brown-Rho:96}, the in-medium vector mesons whose mass
must be much higher than that of quasiquarks were integrated out
and the SUS was then given by the RPA bubble of
quasiquark-quasiantiquark excitations in NJL model. In this paper
where the vector meson plays the key role, it is not clear that we
are not double-counting if we introduce $both$ the vector meson
and the quasiquark in HLS/VM theory. Our point of view here is
that whatever fermionic degrees of freedom are to be implemented
in the theory should be color-singlet and hence the relevant
fermionic degrees of freedom must be baryons. Assuming that the
baryons are always more massive in the heat bath, our results
should then correspond to HLS/VM with the baryons integrated out.
Though our results clearly indicate the dual nature of the
fermionic RPA bubble~\cite{Brown-Rho:96} and the one-loop HLS/VM,
how to consistently implement fermions in a calculation of the
type we considered here is still an open issue. We intend to
return to this in a later publication.

So far we have not addressed the properties of the quantities we
have studied in this paper away from the critical point $T_c$. The
``intrinsic dependence" crucial in our formalism is a difficult
problem to solve away from the $T=0$ and $T=T_c$ points. We have
not yet formulated how to go about this problem. Confrontation
with future lattice data as well as with RHIC data will obviously
require these properties to be worked out. In the case of density,
this problem has been approached from a different perspective in
\cite{LPRVchi}. It is plausible that a similar approach can be
developed for the temperature case.

A more comprehensive discussion of the materials covered in this
paper as well as other issues of HLS-VM in hot bath near chiral
restoration  will be discussed in a future
publication~\cite{HaradaSasaki:prep}.

In the present analysis, we used the bare HLS Largangian with
Lorentz invariance since the Wilsonian matching between the HLS
and the underlying QCD showed that the Lorentz violating effects
to the bare parameters of the HLS Lagrangian are small, as we
discussed around the end of section~\ref{sec:HLS} and
Appendix~\ref{app:WM}. The details of the inclusion of such small
corrections at the bare level and quantum effects based on such a
Lagrangian will be presented in future publications.

\subsection*{Acknowledgments}
\indent\indent We acknowledge useful discussions with Gerry Brown
and Bengt Friman. The work of MH was supported in part by the
Brain Pool program (\#012-1-44) provided by the Korean Federation
of Science and Technology Societies and USDOE Grant
\#DE-FG02-88ER40388.  He would like to thank Gerry Brown for his
hospitality during the stay at SUNY at Stony Brook, and Dong-Pil
Min for his hospitality during the stay at Seoul National
University where part of this work was done. The work of YK was
partially supported by the Brain Korea 21 project of the Ministry
of Education, by the KOSEF Grant No. R01-1999-000-00017-0, and by
the U.S. NSF Grant No. INT-9730847. He is very grateful to
Kuniharu Kubodera and Fred Myhrer for their hospitality during his
stay at University of South Carolina where part of this paper was
written. The work of MR was supported by the Humboldt Foundation
while he was spending three months in the Spring 2001 at the
Theory Group, GSI (Darmstadt, Germany). He would like to thank GSI
for the hospitality and the Humboldt Foundation for the support.

\newpage

\appendix

\begin{flushleft}
\Large\bf Appendices
\end{flushleft}

\section{Wilsonian Matching and Intrisic Temperature Dependences}
\label{app:WM}

The Wilsonian matching was originally proposed at $T=0$~\cite{HY:WM} to
determine the bare parameters of the HLS by matching the HLS to
the underlying QCD.
In Ref.~\cite{HaradaSasaki} the Wilsonian matching was extended to
non-zero temperature
and it was shown that the parameters of the HLS Lagrangian have the
{\it intrinsic temperature dependences}.
In this appendix we first briefly review
the Wilsonian matching proposed in Ref.~\cite{HY:WM} at $T=0$
to determine the bare parameters of the HLS Lagrangian
by matching the HLS with the underlying QCD.
(For details, see Ref.~\cite{HY:PR}.)
Then, we extend the Wilsonian matching
to the analysis at non-zero temperature
to
determine the
intrinsic temperature dependences of the bare
parameters of the (bare) HLS Lagrangian needed in the present
paper.

\subsection{Wilsonian matching conditions at $T=0$}
\label{ssec:WMCT0}

The Wilsonian matching proposed in Ref.~\cite{HY:WM}
is done by matching
the axial-vector and vector current correlators derived from the
HLS with those by the operator product expansion (OPE) in
QCD at the matching scale $\Lambda$.~\footnote{%
  For the validity of the expansion in the HLS the
  matching scale $\Lambda$ must be smaller than the chiral symmetry
  breaking scale $\Lambda_\chi$.
}
The axial-vector and vector current correlators in the OPE
up until ${\cal O}(1/Q^6)$
at $T=0$ are expressed as~\cite{SVZ}
\begin{eqnarray}
\Pi_A^{\rm(QCD)}(Q^2) &=& \frac{1}{8\pi^2}
\left( \frac{N_c}{3} \right)
\Biggl[
  - \left(
      1 +  \frac{3(N_c^2-1)}{8N_c}\, \frac{\alpha_s}{\pi}
  \right) \ln \frac{Q^2}{\mu^2}
\nonumber\\
&& \qquad
  {}+ \frac{\pi^2}{N_c}
    \frac{
      \left\langle
        \frac{\alpha_s}{\pi} G_{\mu\nu} G^{\mu\nu}
      \right\rangle
    }{ Q^4 }
  {}+ \frac{\pi^3}{N_c} \frac{96(N_c^2-1)}{N_c^2}
    \left( \frac{1}{2} + \frac{1}{3N_c} \right)
    \frac{\alpha_s \left\langle \bar{q} q \right\rangle^2}{Q^6}
\Biggr]
\ ,
\label{Pi A OPE}
\\
\Pi_V^{\rm(QCD)}(Q^2) &=& \frac{1}{8\pi^2}
\left( \frac{N_c}{3} \right)
\Biggl[
  - \left(
      1 +  \frac{3(N_c^2-1)}{8N_c}\, \frac{\alpha_s}{\pi}
  \right) \ln \frac{Q^2}{\mu^2}
\nonumber\\
&& \qquad
  {}+ \frac{\pi^2}{N_c}
    \frac{
      \left\langle
        \frac{\alpha_s}{\pi} G_{\mu\nu} G^{\mu\nu}
      \right\rangle
    }{ Q^4 }
  {}- \frac{\pi^3}{N_c} \frac{96(N_c^2-1)}{N_c^2}
    \left( \frac{1}{2} - \frac{1}{3N_c} \right)
    \frac{\alpha_s \left\langle \bar{q} q \right\rangle^2}{Q^6}
\Biggr]
\ ,
\label{Pi V OPE}
\end{eqnarray}
where $\mu$ is the renormalization scale of QCD
and we
wrote the $N_c$-dependences explicitly
(see, e.g., Ref.~\cite{Bardeen-Zakharov}).
In the HLS the same correlators are
well described by the tree contributions with including
${\cal O}(p^4)$ terms
when the momentum is around the matching scale, $Q^2 \sim \Lambda^2$:
\begin{eqnarray}
\Pi_A^{\rm(HLS)}(Q^2) &=&
\frac{F_\pi^2(\Lambda)}{Q^2} - 2 z_2(\Lambda)
\ ,
\label{Pi A HLS}
\\
\Pi_V^{\rm(HLS)}(Q^2) &=&
\frac{
  F_\sigma^2(\Lambda)
}{
  M_\rho^2(\Lambda) + Q^2
}
\left[ 1 - 2 g^2(\Lambda) z_3(\Lambda) \right]
- 2 z_1(\Lambda)
\ ,
\label{Pi V HLS}
\end{eqnarray}
where we defined the bare $\rho$ mass $M_\rho(\Lambda)$ as
\begin{equation}
M_\rho^2(\Lambda) \equiv g^2(\Lambda) F_\sigma^2(\Lambda)
\ .
\label{on-shell cond 5}
\end{equation}

We require that current correlators in the HLS
in Eqs.~(\ref{Pi A HLS}) and (\ref{Pi V HLS})
can be matched with those in QCD in
Eqs.~(\ref{Pi A OPE}) and (\ref{Pi V OPE}).
Of course,
this matching cannot be made for any value of $Q^2$,
since the $Q^2$-dependences of the current correlators
in the HLS are completely
different from those in the OPE:
In the HLS the derivative expansion (in {\it positive} power of $Q$)
is used, and the expressions for
the current correlators are valid in the low energy region.
The OPE, on the other hand, is an asymptotic expansion
(in {\it negative} power of $Q$), and it is
valid in the high energy region.
Since we calculate the current correlators in the HLS including the
first non-leading order [${\cal O}(p^4)$], we expect that we can match
the correlators with those in the OPE up until the first derivative.
Then we obtain the following Wilsonian matching
conditions~\cite{HY:WM,HY:PR}~\footnote{%
  One might think that there appear corrections from $\rho$ and/or
  $\pi$ loops in the left-hand-sides of Eqs.~(\ref{match A}) and
  (\ref{match V}).
  However, such corrections are of higher order in the present
  counting scheme, and thus we neglect them here
  at $Q^2 \sim {\Lambda}^2$.
  In the low-energy scale
  we incorporate the loop effects into the correlators.
  }
\begin{eqnarray}
&&
   \frac{F^2_\pi (\Lambda)}{{\Lambda}^2}
  = \frac{1}{8{\pi}^2} \left( \frac{N_c}{3} \right)
  \Biggl[
    1 +
    \frac{3(N_c^2-1)}{8N_c}\, \frac{\alpha_s}{\pi}
    + \frac{2\pi^2}{N_c}
      \frac{
        \left\langle
          \frac{\alpha_s}{\pi} G_{\mu\nu} G^{\mu\nu}
        \right\rangle
      }{ \Lambda^4 }
\nonumber\\
&& \qquad\qquad\qquad\qquad
    {}+ \frac{288\pi(N_c^2-1)}{N_c^3}
      \left( \frac{1}{2} + \frac{1}{3N_c} \right)
      \frac{\alpha_s \left\langle \bar{q} q \right\rangle^2}
           {\Lambda^6}
  \Biggr]
\ ,
\label{match A}
\\
&&
   \frac{F^2_\sigma (\Lambda)}{{\Lambda}^2}
        \frac{{\Lambda}^4[1 - 2g^2(\Lambda)z_3(\Lambda)]}
         {({M_\rho}^2(\Lambda) + {\Lambda}^2)^2}
\nonumber\\
&& \qquad
= \frac{1}{8\pi^2} \left( \frac{N_c}{3} \right)
  \Biggl[
    1 +
    \frac{3(N_c^2-1)}{8N_c}\, \frac{\alpha_s}{\pi}
    + \frac{2\pi^2}{N_c}
      \frac{
        \left\langle
          \frac{\alpha_s}{\pi} G_{\mu\nu} G^{\mu\nu}
        \right\rangle
      }{ \Lambda^4 }
\nonumber\\
&& \qquad\qquad\qquad\qquad
    {}- \frac{288\pi(N_c^2-1)}{N_c^3}
      \left( \frac{1}{2} - \frac{1}{3N_c} \right)
      \frac{\alpha_s \left\langle \bar{q} q \right\rangle^2}
           {\Lambda^6}
  \Biggr]
\ ,
\label{match V}
\\
&&
   \frac{F^2_\pi (\Lambda)}{{\Lambda}^2} -
            \frac{F^2_\sigma (\Lambda)[1 -
              2g^2(\Lambda)z_3(\Lambda)]}
             {{M_\rho}^2(\Lambda) + {\Lambda}^2} -
            2[z_2(\Lambda) - z_1(\Lambda)]
\nonumber\\
&& \qquad
  =  \frac{4\pi(N_c^2-1)}{N_c^2}
  \frac{\alpha_s \left\langle \bar{q} q \right\rangle^2}{\Lambda^6}
\ .
\label{match z}
\end{eqnarray}
The above three equations (\ref{match z}), (\ref{match A}) and
(\ref{match V}) are the Wilsonian matching conditions
proposed in Ref.~\cite{HY:WM}.
These determine several bare parameters of the HLS without much
ambiguity.  Especially, the first condition (\ref{match A})
determines the ratio $F_\pi(\Lambda)/\Lambda$ directly from QCD.

\subsection{``Intrinsic" temperature dependence of the bare
  parameters}
\label{ssec:ITDBP}

Let us consider the extension of the above matching conditions to
the analysis in hot matter. The quark condensate as well as the
gluon condensate appearing in the right-hand-side (RHS) of the
above matching conditions generally have the temperature
dependence which is converted into the {\it intrinsic temperature
dependence} of bare parameters through the matching
conditions~\cite{HaradaSasaki}. We should note that there is no
longer Lorentz symmetry in hot matter, and the Lorentz non-scalar
operators such as $\bar{q}\gamma_\mu D_\nu q$ may exist in the
form of the current correlators derived by the
OPE~\cite{Hatsuda-Koike-Lee}. Such a contribution may generate
Lorentz symmetry violating effects in the RHS of the above
matching conditions, and accordingly, we may have to use the bare
Lagrangian with the Lorentz symmetry breaking effects included as
in Appendix~A of Ref.~\cite{HKR}. However, we will see that we can
use the Lorentz invariant form of the bare Lagrangian in
Eq.~(\ref{Lagrangian}) near the critical temperature as a good
approximation as follows:

In the RHS of the matching condition in Eq.~(\ref{match A}),
the Lorentz symmetry violating contribution from the
operators such as $\bar{q}\gamma_\mu D_\nu q$
are small compared with the main term of $1 + \frac{\alpha_s}{\pi}$.
This implies that
the Lorentz symmetry breaking effect in the
left-hand-side of Eq.~(\ref{match A}), which is expressed by the
Lorentz symmetry violation in the {\it bare} $\pi$ decay constant,
is also small:
The difference between $F_{\pi,{\rm bare}}^t$ and
$F_{\pi,{\rm bare}}^s$ is small compared with their own values,
or equivalently, the bare $\pi$ velocity defined by
$v_{\pi,{\rm bare}}^2 \equiv
  F_{\pi,{\rm bare}}^s / F_{\pi,{\rm bare}}^t$ is close to one.
As a result we can determine, in a good approximation, the bare
$\pi$ decay constant through the matching condition in
Eq.~(\ref{match A}) with putting possible temperature dependence
on the gluon and quark condensates~\cite{HaradaSasaki}:
\begin{eqnarray}
&&
   \frac{F^2_\pi (\Lambda ;T)}{{\Lambda}^2}
    = \frac{1}{8{\pi}^2}
    \Biggl[
        1 +  \frac{\alpha_s}{\pi}
        + \frac{2\pi^2}{3}
           \frac{\langle \frac{\alpha _s}{\pi}
            G_{\mu \nu}G^{\mu \nu} \rangle_T }
                                {{\Lambda}^4}
       {}+ \pi^3 \,\frac{1408}{27}
                                    \frac{\alpha _s{\langle \bar{q}q
                                         \rangle }^2_T}
                                     {{\Lambda}^6}
                                    \Biggr]
\ ,
\label{eq:WMC A}
\end{eqnarray}
where we took $N_c=3$. We should stress again that, through the
above condition (\ref{eq:WMC A}), the temperature dependence of
the quark and gluon condensates determines the intrinsic
temperature dependence of the {\it bare} $\pi$ decay constant
$F_\pi(\Lambda;T)$.

Next, we consider the intrinsic temperature dependence of other
parameters near the critical temperature. As was shown in
Ref.~\cite{HKR} for the VM in dense matter, the equality between
the vector and axial-vector current correlators in the HLS
requires the following VM conditions for the bare parameters at
the leading order, which should be valid also in hot matter:
\begin{eqnarray}
&&
a_{\rm bare}^t \equiv
  \left(
    \frac{F_{\sigma,{\rm bare}}^t }{ F_{\pi,{\rm bare}}^t }
  \right)^2
  \rightarrow 1
\ ,
\quad
a_{\rm bare}^s \equiv
  \left(
    \frac{F_{\sigma,{\rm bare}}^s }{ F_{\pi,{\rm bare}}^s }
  \right)^2
  \rightarrow 1
\ ,
\label{VM cond a}
\\
&&
g_{T,{\rm bare}} \rightarrow 0 \ , \quad
g_{L,{\rm bare}} \rightarrow 0 \ ,
\qquad
\mbox{for} \ T \rightarrow T_c \ .
\label{VM cond g}
\end{eqnarray}
The VM conditions for the $a$ parameter in Eq.~(\ref{VM cond a})
together with the above result that the Lorentz symmetry violating
effect between $F_{\pi,{\rm bare}}^t$ and $F_{\pi,{\rm bare}}^s$
is small already implies that the effect of Lorentz symmetry
breaking between $F_{\sigma,{\rm bare}}^t$ and $F_{\sigma,{\rm bare}}^s$ is small:
The bare velocity of $\sigma$ (longitudinal $\rho$) defined by
$v_{\sigma,{\rm bare}}^2 =
  F_{\sigma,{\rm bare}}^s / F_{\sigma,{\rm bare}}^t$
is close to one near the critical temperature determined
from the intrinsic temperature dependence.
On the other hand, the ratio
$v_{T,{\rm bare}} \equiv g_{L,{\rm bare}}/g_{T,{\rm bare}}$,
which we call the bare velocity of the transverse $\rho$, 
 cannot be determined through the Wilsonian matching, 
 since the transverse $\rho$ decouples near the critical point  in the 
VM~\cite{HY:VM,HY:PR,HaradaSasaki}.
However, this decoupling nature of the transverse $\rho$
near the critical temperature implies that it becomes
irrelevant to the quantities studied in this paper.
Thus, in the present analysis, we set $v_{T,{\rm bare}}=1$
for simplicity of the calculation, and show how
the transverse $\rho$ decouples from the quantities we
study in this paper.

\section{Hadronic Thermal Corrections}
\label{app:HTC}
 \indent\indent In this appendix we summarize the
hadronic thermal corrections to the two-point functions of
$\overline{\cal A}_\mu$-$\overline{\cal A}_\nu$,
$\overline{V}_\mu$-$\overline{V}_\nu$, $\overline{\cal
V}_\mu$-$\overline{\cal V}_\nu$ and
$\overline{V}_\mu$-$\overline{\cal V}_\nu$.

The four components of the hadronic thermal corrections to the two
point function of $\overline{\cal A}_\mu$-$\overline{\cal A}_\nu$,
$\Pi_\perp$, are expressed as
\begin{eqnarray}
\bar{\Pi}_{\perp}^t(p_0,\bar{p};T)
&=&
  N_f (a-1) \bar{A}_{0}(0,T)
  - N_f a M_\rho^2 \bar{B}_{0}(p_0,\bar{p};M_\rho,0;T)
\nonumber\\
&&
  {}+ N_f \frac{a}{4} \bar{B}^t(p_0,\bar{p};M_\rho,0;T)
\ ,
\label{AA t}
\\
\bar{\Pi}_{\perp}^s(p_0,\bar{p};T)
&=&
  N_f (a-1) \bar{A}_{0}(0,T)
  - N_f a M_\rho^2 \bar{B}_{0}(p_0,\bar{p};M_\rho,0;T)
\nonumber\\
&&
  {}+ N_f \frac{a}{4} \bar{B}^s(p_0,\bar{p};M_\rho,0;T)
\ ,
\label{AA s}
\\
  \bar{\Pi}_{\perp}^L(p_0,\bar{p};T)
&=&
  N_f \frac{a}{4} \bar{B}^L(p_0,\bar{p};M_\rho,0;T)
\ ,
\label{AA L}
\\
\bar{\Pi}_{\perp}^T(p_0,\bar{p};T)
&=&
  N_f \frac{a}{4} \bar{B}^T(p_0,\bar{p};M_\rho,0;T)
\ ,
\label{AA T}
\end{eqnarray}
where the functions $\bar{A}_{0}$, $\bar{B}_{0}$, and so on
are given in Appendix~\ref{app:Fun}.

The two components $\bar{\Pi}^t$ and $\bar{\Pi}^s$ of hadronic
thermal corrections to the two-point functions of
$\overline{V}_\mu$-$\overline{V}_\nu$, $\overline{\cal
V}_\mu$-$\overline{\cal V}_\nu$ and
$\overline{V}_\mu$-$\overline{\cal V}_\nu$ are written as
\begin{eqnarray}
&&
\bar{\Pi}_{V}^t(p_0,\bar{p};T)
=
\bar{\Pi}_{V}^s(p_0,\bar{p};T)
\nonumber\\
&& \
=
\bar{\Pi}_{\parallel}^t(p_0,\bar{p};T)
=
\bar{\Pi}_{\parallel}^s(p_0,\bar{p};T)
\nonumber\\
&& \
=
- \bar{\Pi}_{V\parallel}^t(p_0,\bar{p};T)
=
- \bar{\Pi}_{V\parallel}^s(p_0,\bar{p};T)
\nonumber\\
&& \quad
=
- N_f \frac{1}{4}
  \left[ \bar{A}_{0}(M_\rho;T) + a^2 \bar{A}_{0}(0;T) \right]
- N_f M_\rho^2 \bar{B}_{0}(p_0,\bar{p};M_\rho,M_\rho;T)
\ .
\label{rr vv rv ts}
\end{eqnarray}
Among the remaining components only $\bar{\Pi}_\parallel^L$
is relevant to the present analysis.
This is given by~\footnote{%
  The explicit forms of other components will be listed in
  Ref.~\cite{HaradaSasaki:prep}.
}
\begin{eqnarray}
\bar{\Pi}_{\parallel}^L(p_0,\bar{p};T)
&=&
N_f \frac{1}{8} \bar{B}^L(p_0,\bar{p};M_\rho,M_\rho;T)
+ N_f \frac{(2-a)^2}{8} \bar{B}^L(p_0,\bar{p};0,0;T)
\ .
\label{vv L}
\end{eqnarray}

For obtaining the pion decay constants and velocity in
Section~\ref{sec:PDCV} we need the limit of $p_0 = \bar{p}$ of
$\bar{\Pi}_\perp^t$ and $\bar{\Pi}_\perp^s$ in Eqs.~(\ref{AA t})
and (\ref{AA s}).
With Eq.~(\ref{B0 Bts VM limits}),
$\bar{\Pi}_\perp^t$ and $\bar{\Pi}_\perp^s$ reduce to the
following forms in the limit $M_\rho \rightarrow0$ and
$a\rightarrow1$:
\begin{eqnarray}
&&
\bar{\Pi}_\perp^t(p_0=\bar{p}+i\epsilon,\bar{p};T)
\ \mathop{\longrightarrow}_{M_\rho \rightarrow 0,\, a\rightarrow1} \
- \frac{N_f}{2} \widetilde{J}_{1}^2(0;T)
= - \frac{N_f}{24} T^2
\ ,
\nonumber\\
&&
\bar{\Pi}_\perp^s(p_0=\bar{p}+i\epsilon,\bar{p};T)
\ \mathop{\longrightarrow}_{M_\rho \rightarrow 0,\, a\rightarrow1} \
- \frac{N_f}{2} \widetilde{J}_{1}^2(0;T)
= - \frac{N_f}{24} T^2
\ .
\label{PiA ts Tc app}
\end{eqnarray}

{}In the static--low-momentum limits of the functions listed in
Eq.~(\ref{JB SL}), the $(\bar{\Pi}_\perp^t - \bar{\Pi}_\perp^L)$
appearing in the axial-vector susceptibility
becomes
\begin{eqnarray}
&&
\lim_{\bar{p}\rightarrow0}
\lim_{p_0\rightarrow0}
\left[
  \bar{\Pi}_\perp^t(p_0,\bar{p};T) - \bar{\Pi}_\perp^L(p_0,\bar{p};T)
\right]
\nonumber\\
&& \quad
=
- N_f \widetilde{J}_{1}^2(0;T)
+ N_f a\, \widetilde{J}_{1}^2(M_\rho;T)
- N_f \frac{a}{M_\rho^2}
\left[
  \widetilde{J}_{-1}^2(M_\rho;T)
  - \widetilde{J}_{-1}^2(0;T)
\right]
\ .
\label{PiA tmL SL}
\end{eqnarray}
For the functions appearing in the vector susceptibility
relevant to the present analysis
we have
\begin{eqnarray}
\lim_{\bar{p}\rightarrow0}
\lim_{p_0\rightarrow0}
\left[ \bar{\Pi}_V^t(p_0,\bar{p};T) \right]
&=&
- \frac{N_f}{4} \left[
  2 \widetilde{J}_{-1}^0(M_\rho;T)
  - \widetilde{J}_{1}^2(M_\rho;T)
  + a^2 \, \widetilde{J}_{1}^2(0;T)
\right]
\ ,
\label{Pir t SL}
\\
\lim_{\bar{p}\rightarrow0}
\lim_{p_0\rightarrow0}
\left[ \bar{\Pi}_{\parallel}^L(p_0,\bar{p};T) \right]
&=&
- N_f \frac{1}{4} \left[
  M_\rho^2 \widetilde{J}_{1}^0(M_\rho;T)
  + 2 \widetilde{J}_{1}^2(M_\rho;T)
\right]
\nonumber\\
&&
{}- N_f \frac{(2-a)^2}{2} \widetilde{J}_{1}^2(0;T)
\ .
\label{Piv L SL}
\end{eqnarray}

\section{Functions}
\label{app:Fun}
 \indent\indent In this appendix we list the explicit forms of the
functions that figure in the hadronic thermal corrections,
$\bar{A}_{0}$, $\bar{B}_{0}$ and
$\bar{B}^{\mu\nu}$ in various limits relevant to the
present analysis.

The function
$\bar{A}_{0}(M;T)$ is
expressed as
\begin{eqnarray}
&&
\bar{A}_{0}(M;T) = \widetilde{J}_{1}^2(M;T)
\ ,
\label{A0 J}
\end{eqnarray}
where $\tilde{J}_{1}^2(M;T)$ is defined by
\begin{eqnarray}
&&
\widetilde{J}_{l}^n(M;T)
= \int \frac{d^3\vec{k}}{(2\pi)^3}
\frac{1}{ e^{\omega(\vec{k};M)/T} - 1 }
\frac{ \vert \vec{k} \vert^{n-2} }{[ \omega(\vec{k};M) ]^l }
\ ,
\end{eqnarray}
with $l$ and $n$ being integers and $\omega(\vec{k};M) \equiv
\sqrt{ M^2 + \vert\vec{k}\vert^2 }$. In the massless limit $M=0$,
the above integration can be performed analytically.
Here we list the result relevant to the present analysis:
\begin{eqnarray}
&&
\widetilde{J}_{1}^2(0;T) =
\widetilde{J}_{-1}^0(0;T) = \frac{1}{12} T^2 \ .
\end{eqnarray}

It is convenient to decompose $\bar{B}^{\mu\nu}$
into four components as done for $\Pi_\perp^{\mu\nu}$
in Eq.~(\ref{Pi perp decomp}):
\begin{equation}
\bar{B}^{\mu\nu}
 =u^\mu u^\nu \bar{B}^t +
   (g^{\mu\nu}-u^\mu u^\nu) \bar{B}^s +
   P_L^{\mu\nu} \bar{B}^L + P_T^{\mu\nu}\bar{B}^T
\ .
\label{Bmn decomp}
\end{equation}
We note here that, by explicit computations, the following
relations are satisfied:
\begin{eqnarray}
&&
\bar{B}^t(p_0,\bar{p};M,M;T) =
\bar{B}^s(p_0,\bar{p};M,M;T) =
- 2 \bar{A}_{0}(M;T) = -2 \widetilde{J}_{1}^2(M;T)
\ .
\label{Bts rel}
\end{eqnarray}

To obtain the pion decay constants and velocity in
Section~\ref{sec:PDCV} we need the limit of $p_0 = \bar{p}$ of the
functions in Eqs.~(\ref{AA t}) and (\ref{AA s}).
As for the functions $M_\rho^2\bar{B}_{0}$,
$\bar{B}^t$ and $\bar{B}^s$ appearing
in Eqs.~(\ref{AA t}) and (\ref{AA s}),
we find
that, in the limit of $M_\rho$ going to zero,
they reduce to
\begin{eqnarray}
&&
M_\rho^2 \bar{B}_0(p_0=\bar{p} + i \epsilon,\bar{p};M_\rho,0;T)
\ \mathop{\longrightarrow}_{M_\rho \rightarrow 0} \
0 \ ,
\nonumber\\
&&
\bar{B}^t(p_0=\bar{p}+ i\epsilon,\bar{p};M_\rho,0;T)
\ \mathop{\longrightarrow}_{M_\rho \rightarrow 0} \
- 2 \widetilde{J}_{1}^2(0;T) = - \frac{1}{6} T^2 \ ,
\nonumber\\
&&
\bar{B}^s(p_0=\bar{p}+ i\epsilon,\bar{p};M_\rho,0;T)
\ \mathop{\longrightarrow}_{M_\rho \rightarrow 0} \
- 2 \widetilde{J}_{1}^2(0;T) = - \frac{1}{6} T^2 \ .
\label{B0 Bts VM limits}
\end{eqnarray}

The static--low-momentum limits of the functions appearing in the
corrections to the axial-vector and vector susceptibility
are summarized as
\begin{eqnarray}
&&
\lim_{\bar{p}\rightarrow0}
\lim_{p_0\rightarrow0}
\left[ M_\rho^2 \bar{B}_{0}(p_0,\bar{p};M_\rho,0;T) \right]
=
- \widetilde{J}_{1}^2(M_\rho;T) + \widetilde{J}_{1}^2(0;T)
\ ,
\nonumber\\
&&
\lim_{\bar{p}\rightarrow0}
\lim_{p_0\rightarrow0}
\left[
  \bar{B}^t(p_0,\bar{p};M_\rho,0;T)
  - \bar{B}^L(p_0,\bar{p};M_\rho,0;T)
\right]
\nonumber\\
&& \qquad
=
\frac{-4}{M_\rho^2} \left[
  - \widetilde{J}_{-1}^2(M_\rho;T)
  + \widetilde{J}_{-1}^2(0;T)
\right]
\ ,
\nonumber\\
&&
\lim_{\bar{p}\rightarrow0}
\lim_{p_0\rightarrow0}
\left[ M_\rho^2 \bar{B}_{0}(p_0,\bar{p};M_\rho,M_\rho;T) \right]
=
\frac{1}{2}
\left[
  \widetilde{J}_{-1}^0(M_\rho;T)
  - \widetilde{J}_{1}^2(M_\rho;T)
\right]
\ ,
\nonumber\\
&&
\lim_{\bar{p}\rightarrow0}
\lim_{p_0\rightarrow0}
\left[ \bar{B}^L(p_0,\bar{p};M_\rho,M_\rho;T) \right]
=
- 2 M_\rho^2 \widetilde{J}_{1}^0(M_\rho;T)
- 4 \widetilde{J}_{1}^2(M_\rho;T)
\ ,
\nonumber\\
&&
\lim_{\bar{p}\rightarrow0}
\lim_{p_0\rightarrow0}
\left[ \bar{B}^L(p_0,\bar{p};0,0;T) \right]
=
- 4 \widetilde{J}_{1}^2(0;T)
\ .
\label{JB SL}
\end{eqnarray}


\end{document}